\RequirePackage[dvipsnames]{xcolor}
\documentclass[journal,twoside,web]{ieeecolor}

\usepackage{etoolbox}
\makeatletter
\@ifundefined{color@begingroup}%
{\let\color@begingroup\relax
\let\color@endgroup\relax}{}%
\def\fix@ieeecolor@hbox#1{%
\hbox{\color@begingroup#1\color@endgroup}}
\patchcmd\@makecaption{\hbox}{\fix@ieeecolor@hbox}{}{\FAILED}
\patchcmd\@makecaption{\hbox}{\fix@ieeecolor@hbox}{}{\FAILED}

\usepackage{tmi}
\usepackage{cite}
\usepackage{amsmath,amssymb,amsfonts}
\usepackage{bm}
\usepackage{cases}
\usepackage{graphicx}
\usepackage{textcomp}
\usepackage{algpseudocode,algorithm}
\makeatletter
\usepackage{xspace}
\usepackage{mathtools}
\usepackage{tabularray}
\usepackage{multirow}
\usepackage{booktabs}

\def\@onedot{\ifx\@let@token.\else.\null\fi\xspace}
\DeclareRobustCommand\onedot{\futurelet\@let@token\@onedot}
\newcommand{\eqnref}[1]{Eq\onedot~\eqref{#1}}  
\def\BibTeX{{\rm B\kern-.05em{\sc i\kern-.025em b}\kern-.08em
    T\kern-.1667em\lower.7ex\hbox{E}\kern-.125emX}}

\makeatletter
\let\NAT@parse\undefined
\makeatother
\usepackage[colorlinks,
            linkcolor=PineGreen,
            anchorcolor=blue,
            urlcolor=PineGreen,
            citecolor=PineGreen]{hyperref}

\usepackage{array}

\DeclareMathOperator{\st}{s.t.}     
\algnewcommand{\LineComment}[1]{\State \(\triangleright\) #1}

\begin{document}
\title{DPER: Diffusion Prior Driven Neural Representation for Limited Angle and Sparse View CT Reconstruction}
\author{Chenhe Du, Xiyue Lin, Qing Wu, Xuanyu Tian, Ying Su, Zhe Luo, \\Rui Zheng, Yang Chen,~\IEEEmembership{Senior Member, IEEE}, Hongjiang Wei, \\S. Kevin Zhou,~\IEEEmembership{Fellow, IEEE}, Jingyi Yu,~\IEEEmembership{Fellow, IEEE} and Yuyao Zhang, \IEEEmembership{Member, IEEE}
    \thanks{C. Du and X. Lin contributed equally to this manuscript.}
    \thanks{This work was supported by the National Natural Science Foundation of China under Grants 62071299 and 12074258. \textit{(Corresponding author: Yuyao Zhang.)}}
    \thanks{C. Du, X. Lin, Q. Wu, X. Tian, R. Zheng, J. Yu and Y. Zhang are with the School of Information Science and Technology, ShanghaiTech University, Shanghai, China (e-mail: duchenhe@shanghaitech.edu.cn; linxy2022@shanghaitech.edu.cn; wuqing@shanghaitech.edu.cn; tianxy@shanghaitech.edu.cn; zhengrui@shanghaitech.edu.cn; yujingyi@shanghaitech.edu.cn; zhangyy8@shanghaitech.edu.cn).}
    \thanks{Y. Su, Z. Luo are with the Department of Critical Care Medicine, Zhongshan Hospital, Fudan University, Shanghai, China (e-mail: su.ying@zs-hospital.sh.cn; luo.zhe@zs-hospital.sh.cn).}
    \thanks{Y. Chen is with the Laboratory of Image Science and Technology, the School of Computer Science and Engineering, and the Key Laboratory of New Generation Artificial Intelligence Technology and Its Interdisciplinary Applications (Southeast University), Ministry of Education, Nanjing 210096, China. (e-mail: chenyang.list@seu.edu.cn).}
    \thanks{H. Wei is with the School of Biomedical Engineering and Institute of Medical Robotics, Shanghai Jiao Tong University, Shanghai, China (e-mail: hongjiang.wei@sjtu.edu.cn).}
    \thanks{S. Kevin Zhou is with the Center for Medical Imaging, Robotics, Analytic Computing and Learning (MIRACLE), School of Biomedical Engineering, Suzhou Institute for Advanced Research, University of Science and Technology of China, Suzhou 215123, China, and also with the Key Laboratory of Intelligent Information Processing of Chinese Academy of Sciences (CAS), Institute of Computing Technology, CAS, Beijing 100190, China (e-mail: skevinzhou@ustc.edu.cn).}
}
\maketitle
\begin{abstract}
    Limited-angle and sparse-view computed tomography (LACT and SVCT) are crucial for expanding the scope of X-ray CT applications. However, they face challenges due to incomplete data acquisition, resulting in diverse artifacts in the reconstructed CT images.
    Emerging implicit neural representation (INR) techniques, such as NeRF, NeAT, and NeRP, have shown promise in under-determined CT imaging reconstruction tasks.
    However, the unsupervised nature of INR architecture imposes limited constraints on the solution space, particularly for the highly ill-posed reconstruction task posed by LACT and ultra-SVCT.
    In this study, we introduce the Diffusion Prior Driven Neural Representation (DPER), an advanced unsupervised framework designed to address the exceptionally ill-posed CT reconstruction inverse problems.
    DPER adopts the Half Quadratic Splitting (HQS) algorithm to decompose the inverse problem into data fidelity and distribution prior sub-problems. The two sub-problems are respectively addressed by INR reconstruction scheme and pre-trained score-based diffusion model.
    This combination first injects the implicit image local consistency prior from INR. Additionally, it effectively augments the feasibility of the solution space for the inverse problem through the generative diffusion model, resulting in increased stability and precision in the solutions.
        {We conduct comprehensive experiments to evaluate the performance of DPER on LACT and ultra-SVCT reconstruction with two public datasets (AAPM and LIDC), an in-house clinical COVID-19 dataset and a public raw projection dataset created by Mayo Clinic.}
    The results show that our method outperforms the state-of-the-art reconstruction methods on in-domain datasets, while achieving significant performance improvements on out-of-domain (OOD) datasets.
\end{abstract}

\section{Introduction}
\label{sec:introduction}
\IEEEPARstart{X}{-ray} computed tomography (CT) has achieved extensive applications in medical diagnosis, security checks and industrial non-destructive detection \cite{wang2008outlook}. Traditional reconstruction methods, such as Filtered Back Projection (FBP), excel in generating high-quality CT images when complete projection data is available. CT reconstruction from incomplete projection data is one of the key challenges of X-ray CT imaging. The projection data acquired by sparse-view (SV) and limited-angle (LA) sampling are typically incomplete, thus leading to highly ill-posed inverse reconstruction problems. The SVCT effectively mitigates patient exposure to X-ray radiation by minimizing the count of projection views. While the LACT faces limitations in acquiring full-view projections due to CT scanner constraints, such as C-arm imaging systems. Traditional reconstruction leads to significant artifacts in both cases. According to the center-slice theorem, the former involves frequency domain interpolation, while the latter requires extrapolation. Due to more severe data gaps, LACT is harder to solve than SVCT, even with a comparable number of projection views.

\par To improve the image reconstruction performance, numerous model-based iterative reconstruction (MBIR) methods \cite{katsura2012model,liu2014model,andersen1984simultaneous,gordon1970algebraic} have been investigated for solving inverse problems. MBIR problems are often expressed as an optimization challenge using weighted least squares, composed of both a data consistency term and a regularization term.
    {Typical regularization terms include total variation (TV) \cite{liu2012adaptive}, anisotropic TV (ATV) \cite{chen2013limited}, dictionary learning~\cite{xu2012low,wu2018low,chen2014artifact}, and low-rank~\cite{wu2018non}.
        While these methods have demonstrated effectiveness, several limitations impede their widespread application. For example, TV tends to over-smooth image structures, whereas dictionary learning suffers from redundancy and high computational memory requirements.
        To address these issues, researchers have proposed several novel solutions. For instance, $\ell _{0}$-norm regularization~\cite{sun2014image,wu2018low,xu2020limited} has been introduced to mitigate the image structure smoothing inherent in TV methods. Additionally, the convolutional analysis operator learning (CAOL) approach \cite{chun2019convolutional,chun2020momentum} has been suggested as a potential solution to the high memory cost associated with dictionary learning.}
    {Despite the improvement in CT reconstructions, MBIR methods still struggled with removing severe artifacts (\textit{i.e.}, the artifacts in LACT and ultra-SVCT) and suffer from defects like sensitive hyper-parameters and hand-crafted priors.}

\par Benefiting from data-driven priors, supervised deep learning (DL) approaches \cite{wang2020deep} yield state-of-the-art (SOTA) performance for the ill-posed CT reconstruction. For the SVCT task, convolutional neural networks (CNNs) are mainstream architectures \cite{jin2017deep, han2018framing, zhang2018sparse, lee2018deep, li2019learning}. For example, Jin et al. \cite{jin2017deep} proposed FBPConvNet, the first CNN-based SVCT model, which trains a U-Net to transform low-quality images into high-quality images.
    {In addition to direct image-to-image translation methods, combining supervised neural networks with the MBIR method is a more effective scheme~\cite{huang2019data,zhou2021limited,zhang2020metainv,hu2022dior,pan2024iterative}.}
For instance, DCAR \cite{huang2019data} performs SART, an iterative optimization CT reconstruction method, on output images of a well-trained U-Net to further improve data consistency with the SV projections.
    {Adler and Öktem ~\cite{adler2018learned} proposed an approach that unrolls the proximal primal-dual optimization algorithm for tomographic reconstruction, wherein they replace the proximal operators with CNNs.
        Chen et al.~\cite{chen2018learn} and Zhang et al.~\cite{zhang2022learn++} introduced CNNs to learn the gradient of regularization terms in the gradient descent algorithm.
        In addition to image domain post-processing methods, Wu et al. \cite{wu2021drone} introduced a dual-domain method that extracts deep features from both the image acquisition and reconstruction domains, thereby leveraging the strengths of compressed sensing and deep learning techniques.}
The combination of the model-based method and CNNs significantly improves the reconstruction performance. However, these supervised DL methods for SVCT and LACT face two common challenges: (1) The performance of supervised learning methods highly depends on the data distribution of the training dataset (\textit{i.e.}, a large-scale training dataset that covers more types of variations generally provides better performance). (2) Generalization of supervised models is susceptible to varying acquisition protocols (\textit{e.g.}, different SNR levels, projection views, and X-ray beam types), limiting their robustness in clinical applications.

\par Implicit neural representation (INR) is a novel unsupervised DL scheme that demonstrated a significant potential for the SVCT problem \cite{wu2022self, wu2022joint, ruckert2022neat, zha2022naf, zang2021intratomo, sun2021coil, reed2021dynamic, shen2022nerp}. Technically, INR trains a multi-layer perceptron (MLP) to represent the reconstructed CT image as a continuous function that maps the spatial coordinates to the image intensities.
Another method with similar functionality is the deep image prior (DIP) \cite{Ulyanov_2018_CVPR}, which has been applied in CT, PET, and MR image reconstruction \cite{yoo2021time,gong2018pet,baguer2020computed}. DIP operates under the assumption that the neural network architecture inherently provides a regularization effect, and the network parameters are estimated by directly fitting the measurement data without requiring separate training. Compared to DIP, INR introduces spatial coordinate embedding to achieve a continuous representation of the reconstructed image intensities. This approach enhances INR's capability to capture high-frequency image details and has been proven more practical for incorporating various imaging forward models.
For instance, in SVCT reconstruction,  the MLP is optimized by incorporating the CT imaging forward model (\textit{e.g.}, Radon transform) to approximate the SV projection data. The neural representation ensures inherent local image consistency, providing robust implicit regularization to constrain the solution space for the under-determined inverse problem \cite{rahaman2019spectral}.
Nonetheless, this regularization may be less effective in scenarios with more severe data gaps, such as ultra-SVCT (\textit{e.g.}, fewer than 30 projection views) and LACT reconstruction.
As a result, existing INR-based methods hardly produce satisfactory performance in the related tasks.

Recently, score-based diffusion models have emerged as state-of-the-art generative models while also providing excellent generative priors to solve inverse problems in an unsupervised manner.
Score-based diffusion models define a forward process that gradually perturbs the data with noise that transforms the data distribution into Gaussian.
Through the reverse process, the diffusion models learn the data distribution by matching the gradient of the log density.
Pre-trained diffusion models can sample a prior from estimated posterior probabilities when given partial or corrupted measurements.
The prior effectively constrains the solution space for challenging inverse problems.
In medical imaging, diffusion model-based approaches have shown significant advancements in SVCT, LACT, and compressed sensing MRI \cite{song2021solving, chung2022improving, chung2023solving,he2023iterative, peng2022towards,wu2024multi,zhang2024wavelet}.
Moreover, it has been shown that a robust prior is crucial for achieving satisfactory results in challenging scenarios with INR \cite{shen2022nerp}.
Inspired by these advancements, we propose integrating diffusion models as an implicit image prior within the INR optimization process to tackle highly ill-posed inverse problems.

In this work, we propose \textit{\textbf{DPER:} } \textit{\textbf{D}iffusion \textbf{P}rior driven n\textbf{E}ural \textbf{R}epresentation}, an effective unsupervised framework for solving highly ill-posed inverse problems in CT reconstruction, including ultra-SVCT and LACT.
To address the ill-posed nature of CT reconstruction, we integrate score-based diffusion models as a regularizer in the INR optimization to constrain the solution space effectively.
Following plug-and-play (PnP) \cite{venkatakrishnan2013plug,kamilov2023plug} reconstruction methods, we decouple the data fidelity and data prior terms in inverse problem solvers, formulating them as two subproblems optimized alternatively.
Specifically, in the data-fidelity subproblem, we propose a physical model-driven INR-based method. This approach explicitly models the image acquisition process during the fitting of measurement data. As a result, it leverages the advantages of both the physical model and the neural representation continuity prior, resulting in a more comprehensive and accurate representation of the reconstructed images compared to traditional MBIR methods~\cite{shen2022nerp,wu2022self,sun2021coil}.
Regarding data prior, the score-based diffusion model samples a prior image by estimating posterior probabilities based on the reconstruction result from INR.
Through alternating optimizations, DPER ensures that the reconstruction results stay within the target manifold of clean CT images, while keeping exceptional data consistency.
To evaluate the effectiveness of the proposed method, we conduct experiments on SVCT and LACT reconstruction on public AAPM and LIDC datasets, an in-house clinical COVID-19 dataset and a public raw projection dataset.
The main contributions can be summarized as follows:
\begin{enumerate}
    \item {We propose DPER, a novel unsupervised method that fully combines the respective advantages of diffusion modeling and INR, which can effectively solve ultra-SVCT and LACT reconstruction challenging problems.}

    \item {We provide a new perspective for solving the medical imaging inverse problem via diffusion models.
          Innovatively, DPER, integrates diffusion models and INR as operators within a plug-and-play (PnP) framework for image restoration. This alternating iterative optimization process strictly adheres to the PnP framework, theoretically ensuring convergence to the optimal reconstruction result.}
    \item {We provide a new insight of how to add data-driven regularizers to the INR frameworks.
          Our framework successfully and efficiently integrates the state-of-the-art diffusion model as a data prior to INR, creating the potential for broader applications of the INR framework.}
\end{enumerate}

{We conduct comprehensive evaluation experiments, demonstrating that the proposed model significantly outperforms the state-of-the-art DiffusionMBIR \cite{chung2023solving} and MCG \cite{chung2022improving} in both the LACT and ultra-SVCT tasks. Notably, our model achieves an impressive +2 dB improvement in PSNR for the LACT task with a 90° scanning range and an outstanding nearly +5 dB improvement for the LACT task with the same setting on OOD data and real projection data.}

\section{Background}
\label{sec:background}
\subsection{Problem Formulation}
\par The CT acquisition process is typically formulated as a linear forward model:
\begin{equation}
    \mathbf{y}=\mathbf{A}\mathbf{x}+\mathbf{n},
    \label{CT forward model}
\end{equation}
where $\mathbf{y}\in \mathbb{R}^m$ is the measurement data, $\mathbf{A}\in\mathbb{R}^{m\times n}$ is the CT imaging forward model (\textit{e.g.}, Radon transform), $\mathbf{x}\in\mathbb{R}^n$ is the underlying CT image, and $\mathbf{n}\in\mathbb{R}^m$ is the system noise. We seek to reconstruct the unknown CT image $\mathbf{x}$ from the observed measurement $\mathbf{y}$.
\par Under SV and LA CT acquisition settings, the dimension of the measurement $\mathbf{y}$ is often much lower than that of the image $\mathbf{x}$ (\textit{i.e.}, $m\ll n$), resulting in the CT reconstruction task being highly ill-posed without further assumptions. {This problem is often formulated as a maximum-a-posteriori (MAP) estimation \cite{poor2013introduction}, with the objective of maximizing the posterior probability:
        \begin{align}
            \hat{\mathbf{x}} & =\underset{\mathbf{x}}{\mathrm{arg}\max}\ p\left( \mathbf{x}\mid \mathbf{y} \right)                                       \\
                             & =\underset{\mathbf{x}}{\mathrm{arg}\min}-\log p\left( \mathbf{y}\mid \mathbf{x} \right) -\log p\left( \mathbf{x} \right),
            \label{MAP}
        \end{align}
        where the conditional probability $p\left( \mathbf{y} \mid \mathbf{x} \right)$ models the CT forward imaging process, and the prior distribution $p\left( \mathbf{x} \right)$ describes the probability distribution of the prior manifold. The solution to the MAP problem in Eq.~\eqref{MAP} is generally formulated as a regularized inversion~\cite{chan2016plug}, which can be expressed as follows:
    }
\begin{equation}
    \hat{\mathbf{x}}=\underset{\mathbf{x}}{\mathrm{arg}\min}\left\| \mathbf{y}-\mathbf{A}\mathbf{x} \right\| ^2+\lambda\cdot\mathcal{R} \left( \mathbf{x} \right),
    \label{MBIR optimization equation}
\end{equation}
where $\hat{\mathbf{x}}$ denotes the desired CT image to be solved. $\left\| \mathbf{y}-\mathbf{A}\mathbf{x} \right\|^2$ represents a data fidelity term that measures the similarity between the real measurement $\mathbf{y}$ and the estimated measurement $\mathbf{Ax}$. $\mathcal{R} \left( \mathbf{x} \right) $ is a regularization term that imposes certain explicit prior knowledge (\textit{e.g.}, local image consistency~\cite{rudin1992nonlinear}) to constrain the solution space, and $\lambda>0$ is a hyper-parameter that controls the contribution of the regularization term. With the additional regularization, the under-determined linear inverse problem achieves optimized solutions.

\subsection{Implicit Neural Representation}
\par INR is a novel unsupervised framework for solving ill-posed inverse imaging problems. Given an incomplete CT measurement $\mathbf{y}$, INR represents its corresponding CT images $\mathbf{x}$ as a continuous function:
\begin{equation}
    f:\mathbf{p}=(x,y)\in\mathbb{R}^2\rightarrow I\in\mathbb{R},
\end{equation}
where $\mathbf{p}$ denotes any intensity coordinates and $I$ is the image intensity at that position. The function is very complex and intractable. INR thus uses an MLP network $\mathcal{F}_\mathbf{\Phi}$ to fit it through minimizing the following objective:
\begin{equation}
    \mathbf{\Phi}^* = \arg\min_\mathbf{\Phi}\mathcal{L}(\mathbf{A}\mathcal{F}_\mathbf{\Phi}, \mathbf{y}),
\end{equation}
where $\mathcal{L}$ is a metric that computes the distance between the predicted measurement $\mathbf{A}\mathcal{F}_\mathbf{\Phi}$ and real measurement $\mathbf{y}$. $\mathbf{A}$ is the physical imaging forward model (\textit{e.g.}, Radon transform for parallel CT), and $\mathbf{\Phi}^*$ denotes the well-trained weights of the MLP network. By feeding all coordinates $\mathbf{p}$ into the MLP network $\mathcal{F}_\mathbf{\Phi}^*$, the CT images $\hat{\mathbf{x}}$ can be estimated.
\par The INR architectures typically consist of an encoding module (\textit{e.g.}, Fourier encoding \cite{tancik2020fourier} and hash encoding \cite{muller2022instant}) and an MLP network. The encoding module transforms low-dimensional coordinates into high-dimensional embeddings, significantly enhancing the MLP network's ability to fit high-frequency image content. Due to the MLP network's inherent learning bias towards low-frequency signals \cite{rahaman2019spectral} and the integration of the CT forward model, the INR imaging framework has demonstrated significant potential in addressing the SVCT reconstruction problem \cite{wu2022self, wu2022joint, ruckert2022neat, zha2022naf, zang2021intratomo, sun2021coil, reed2021dynamic, shen2022nerp}.

\subsection{Score-Based Diffusion Model}
\subsubsection{Background} The score-based diffusion model is a generative model that progressively transforms pure noise  into the target data distribution through a reverse diffusion process guided by the score function.
It synthesizes samples by gradually reversing the stochastic process, introducing controlled noise while considering data distribution gradients.
The diffusion process $\{\mathbf{x}(t)\}^{T}_{t=0}$ is indexed by a continuous time variable $t \in [0,T]$, such that $\mathbf{x}(0) \sim p_{0}$ and $\mathbf{x}(T) \sim p_{T}$, where $p_{0}$ is the data distribution and $p_{T}$ is the prior distribution (\textit{e.g.}, the Gaussian distribution). The forward diffusion process can be denoted by the solution of the forward stochastic differential equation (SDE):
\begin{equation}
    \mathrm{d}\mathbf{x}=\mathbf{f}(\mathbf{x},t)\mathrm{d}t+g(t)\mathrm{d}\mathbf{w},
    \label{SDE forward process}
\end{equation}
where $\mathbf{w}$ is the standard Wiener process, $\mathbf{f}(\cdot,t)$ and $g(\cdot)$ are the drift and diffusion coefficient of $\mathbf{x}(t)$. By starting from the $\mathbf{x}(T)$, the reverse diffusion process can obtain samples $\mathbf{x}(0)$ solving reverse-time SDE of \eqnref{SDE forward process} as:
\begin{equation}
    \mathrm{d}\mathbf{x}=[\mathbf{f}(\mathbf{x},t) - g(t)^{2} \nabla_{\mathbf{x}}\log{p_{t}(\mathbf{x})}]\mathrm{d}t+g(t)\mathrm{d}\bar{\mathbf{w}},
    \label{SDE reverse process}
\end{equation}
where $\bar{\mathbf{w}}$ is a standard Wiener process that time flows backwards from $T$ to 0. The score function $\nabla_{\mathbf{x}}\log{p_{t}(\mathbf{x})}$ can be estimated by training a time-dependent score-based model $\mathbf{s}_{\bm{\theta}}(\mathbf{x},t)$ by
\begin{equation}
    \begin{aligned}
        \bm{\theta}^{*} = \underset{\bm{\theta}}{\mathrm{arg}\min}\mathbb{E}_{t}
        \Big \{
        \lambda(t)\mathbb{E}_{\mathbf{x}(0)}\mathbb{E}_{\mathbf{x}(t)|\mathbf{x}(0)}[\|\mathbf{s}_{\bm{\theta}}(\mathbf{x},t) \\
        - \nabla_{\mathbf{x}(t)}\log{p_{0t}(\mathbf{x}(t) | \mathbf{x}(0))}\|_{2}^{2}]
        \Big \},
    \end{aligned}
    \label{SDE training equation}
\end{equation}
where $\mathbf{x}(t) \sim p_{0t}(\mathbf{x}(t) | \mathbf{x}(0))$ can be derived from the forward diffusion process, $\lambda(t)$ is the weighting scheme. The trained score model $\mathbf{s}_{\bm{\theta}^{*}}$ can be used in reverse SDE sampling to generate the desired data.

\subsubsection{Diffusion model for inverse problems} Unlike the unconditional diffusion process that samples a prior distribution $p(\mathbf{x})$, when applying the diffusion model to solve an inverse problem, we aim to sample a posterior distribution $p(\mathbf{x}|\mathbf{y})$. Based on Bayes' theorem, the \eqnref{SDE reverse process} can be rewritten as follows:
\begin{equation}
    \mathrm{d}\mathbf{x}=[\mathbf{f}(\mathbf{x},t)-g(t)^2\nabla _{\mathbf{x}}\left( \log p_t(\mathbf{x})+\log p_t(\mathbf{y}| \mathbf{x}) \right) ]\mathrm{d}t+g(t)\mathrm{d}\bar{\mathbf{w}},
\end{equation}
where the posterior is divided into $p_t(\mathbf{x})$ and $p_t(\mathbf{y}|\mathbf{x})$. Consequently, the unconditional pre-trained diffusion models can be harnessed for conditional synthesis, which enables us to regard diffusion model as a prior regularization term for solving inverse problems.
    {However, the noisy likelihood $\log p_t(\mathbf{y}| \mathbf{x})$ is generally intractable. The crux of the problem, therefore, lies in finding methods to approximate it or bypass direct calculation.}

\begin{figure*}[!t]
    \centerline{\includegraphics[width=1\textwidth]{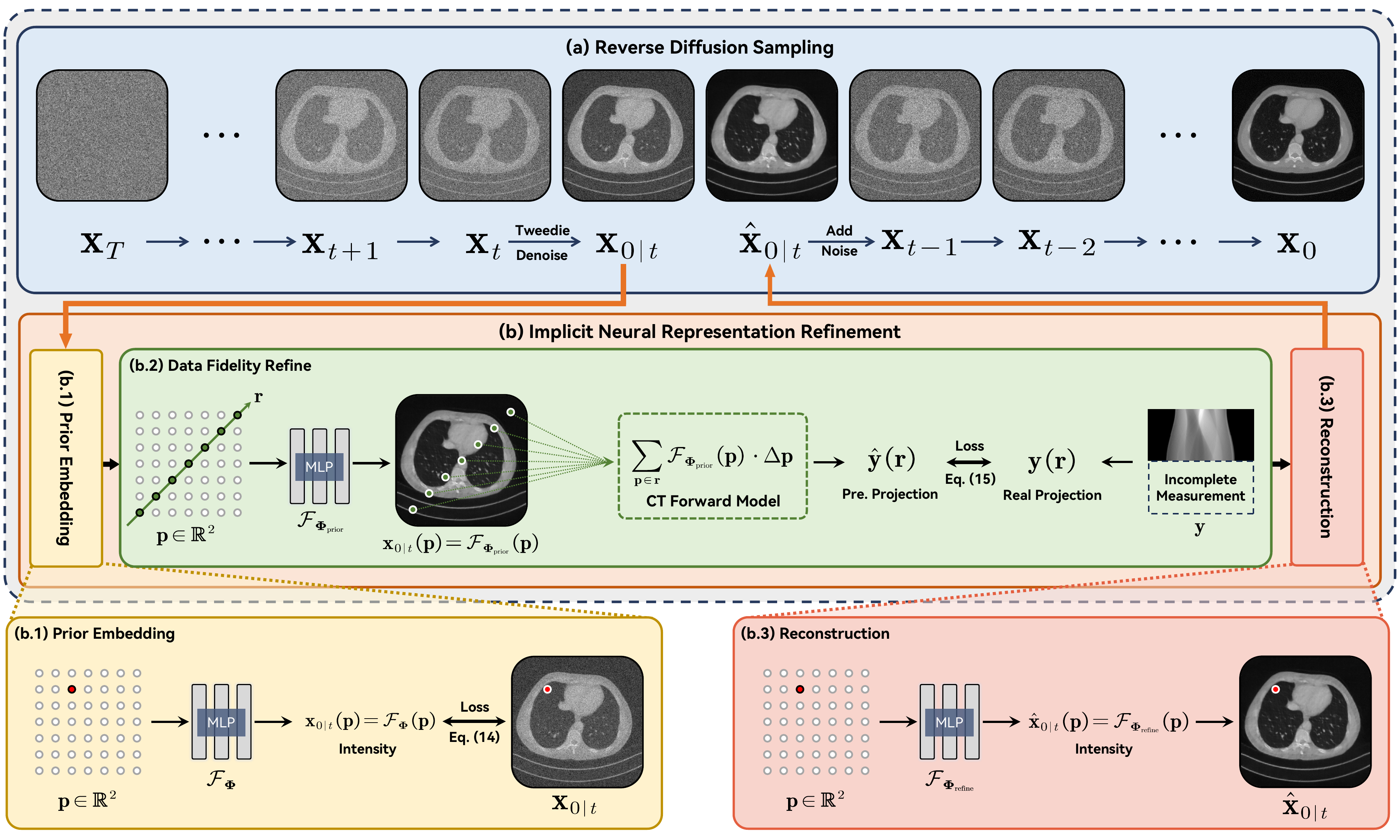}}
    \caption{Overview of the proposed DPER.
    (a) The distribution prior sub-problem is addressed with reverse SDE sampling utilizing a pre-trained score model.
    At the sampling timestep $t$, the noiseless prior $\mathbf{x}_{0|t}$ is generated from $\mathbf{x}_t$ with Tweedie's formula.
    Subsequently, $\mathbf{x}_{0|t}$ is refined with INR by solving the data fidelity subproblem (indicated by the orange arrow) to get $\hat{\mathbf{x}}_{0|t}$.
    Next, $\mathbf{x}_{t-1}$ is derived from $\hat{\mathbf{x}}_{0|t}$ by adding noise, enabling the continuation of the reverse SDE sampling.
    (b) The data fidelity sub-problem is addressed with an INR combing physical forward model of CT.
    First, the prior image $\mathbf{x}_{0|t}$ is embedded into an MLP network $\bm{\mathcal{F}}_{\mathbf{\Phi}}$.
    Then, we simulate the projections using the physical forward model of CT by sampling rays from the implicit representation.
    The parameters of $\bm{\mathcal{F}}_{\mathbf{\Phi}}$ is optimized by minimizing the distance between the predicted projection $\hat{\mathbf{y}}$ and the real projection $\mathbf{y}$.
    Finally, the refined prior image $\hat{\mathbf{x}}_{0|t}$ is generated from well-trained MLP $\bm{\mathcal{F}}_{\hat{\mathbf{\Phi}}}$.
    }
    \label{fig:method_overview}
\end{figure*}

\section{Methodology}
\label{sec:methodology}
\subsection{Overview}
\label{subsec:Overview}
\par LACT and ultra-SVCT present highly under-determined inverse image reconstruction challenges.
The image local consistency prior induced by INR hardly produces dependable solution space in such conditions \cite{shen2022nerp, wu2022self, sun2021coil}.
In this study, we suggest integrating the diffusion model as a generative prior into the INR image reconstruction framework.
The data distribution prior introduced by pre-trained score-based diffusion models serves as image regularization during the INR optimization process.
This combination first preserves the implicit image local consistency prior from INR. Furthermore, it enhances the feasibility of the solution space for the inverse problem through the diffusion model, resulting in increased stability and precision in the solutions.

Revisiting the ill-posed inverse problem as outlined in Eq. (\ref{MBIR optimization equation}), {which can be intuitively understood as finding a balance between the measurement manifold and the prior manifold.} A widely used approach to solve it more efficiently is to decouple the data and prior terms using the split-variable method.
Specifically, we leverage the Half Quadratic Splitting (HQS) \cite{geman1995nonlinear} due to its straightforwardness and ease of implementation. The HQS algorithm first introduces an auxiliary variable $\mathbf{z}$, and the \eqnref{MBIR optimization equation} is equivalent to the constrained optimization problem as:
\begin{equation}
    \left\{ \hat{\mathbf{x}},\hat{\mathbf{z}} \right\} =\underset{\mathbf{x},\mathbf{z}}{\mathrm{arg}\min}\left\| \mathbf{y}-\mathbf{Ax} \right\| ^2+\lambda \cdot \mathcal{R} \left( \mathbf{z} \right),\ \st \ \mathbf{z}=\mathbf{x},
    \label{eq:hqs_1}
\end{equation}
{where the constraint is given by $\mathbf{z}=\mathbf{x}$. The HQS-based penalty method seeks to minimize the following cost function:}
\begin{equation}
    \mathcal{L} _{\mu}\left( \mathbf{x},\mathbf{z} \right) =\left\| \mathbf{y}-\mathbf{Ax} \right\| ^2+\lambda \cdot \mathcal{R} \left( \mathbf{z} \right) +\frac{\mu}{2}\left\| \mathbf{z}-\mathbf{x} \right\| ^2,
    \label{eq:hqs_2}
\end{equation}
where $\frac{\mu}{2}\left\| \mathbf{z}-\mathbf{x} \right\| ^2$ is a penalty term that penalizes the deviation of $\mathbf{z}$ from $\mathbf{x}$. The parameter $\mu$ controls the strength of this penalty. As $\mu$ increases, the penalty term becomes more dominant, leading $\mathbf{x}$ to approach $\mathbf{z}$  more precisely.
The problem can be addressed by iteratively solving the following two subproblems while keeping the rest of the variables fixed:
\begin{subequations}
    \label{eq:m1.hqs}
    \begin{numcases}{}
        \mathbf{x}_{t}=\underset{\mathbf{x}}{\mathrm{arg} \min}\left\| \mathbf{y}-\mathbf{A}\mathbf{x} \right\| ^2+\frac{\mu}{2} \left\| \mathbf{x}-\mathbf{z}_{t-1} \right\| ^2\label{eq:m1.data_sub}\\
        \mathbf{z}_{t}=\underset{\mathbf{z}}{\mathrm{arg} \min}\frac{1}{2(\sqrt{\lambda /\mu})^2}\left\| \mathbf{z}-\mathbf{x}_{t} \right\| ^2+\mathcal{R} \left( \mathbf{z} \right)\label{eq:m1.prior_sub},
    \end{numcases}
\end{subequations}
{since then, the original problem has been transformed into the standard plug-and-play image restoration (IR) framework~\cite{kamilov2023plug}.} Eq. \eqref{eq:m1.data_sub} defines the data-fidelity subproblem, concentrating on the identification of a proximal point in relation to $\mathbf{z}_{t-1}$. {Meanwhile, the prior subproblem in Eq. \eqref{eq:m1.prior_sub} can be interpreted as a denoising problem with additive white Gaussian noise (AWGN) of variance $(\sqrt{\lambda /\mu})^2$~\cite{venkatakrishnan2013plug}.
        To ensure that $\mathbf{x}_{t}$ and $\mathbf{z}_{t}$ in the iterative framework converge to a fixed point (\textit{i.e.}, the optimal solution to the original problem), $\mu$ needs to be gradually increased with iterations towards infinity ($\mu \rightarrow \infty$)\cite{zhang2021plug}. For a detailed discussion on the theoretical convergence of plug-and-play IR, please refer to \cite{chan2016plug}}

Firstly, we describe how to solve the prior subproblem (as Eq. \eqref{eq:m1.prior_sub}) using the score-based diffusion model under the HQS framework in Eq. \eqref{eq:m1.hqs}.
{Leverage the wisdom of PnP, the proximal operator in Eq. \eqref{eq:m1.prior_sub} can be replaced by a general denoiser $\mathcal{D} \left( \cdot ,\sigma^2 \right) :\mathbb{R} ^n\rightarrow \mathbb{R} ^n$ as:
\begin{equation}
    \mathbf{z}_t=\mathcal{D} \left( \mathbf{x}_t,(\sqrt{\lambda /\mu})^2 \right),
\end{equation}
this AWGN denoising problem involves deriving a noise-free $\mathbf{z}_{t}$ from a noisy $\mathbf{x}_{t}$ (\textit{i.e.}, obtain the $\mathbf{x}_{0|t}$), where the noise level is quantified by $\sqrt{\lambda /\mu}$. By introducing the Tweedie's formula \cite{robbins1992empirical}, we can rewrite Eq. \eqref{eq:m1.prior_sub} as:
\begin{align}
    \mathbf{z}_t=\mathbf{x}_{0|t}
     & =\mathcal{D} \left( \mathbf{x}_t,(\sqrt{\lambda /\mu})^2 \right)                                            \\
     & =\mathbb{E} \left[ \mathbf{z}_t|\mathbf{x}_t \right]                                                        \\
     & =\mathbf{x}_t+\left( \sqrt{\lambda /\mu} \right) ^2\nabla \log p\left( \mathbf{x}_t \right)                 \\
     & \approx \mathbf{x}_t+\left( \sqrt{\lambda /\mu} \right) ^2\mathbf{s}_{\theta}\left( \mathbf{x}_t,t \right).
\end{align}}
In alignment with the approach adopted in \cite{zhu2023denoising}, the noise schedule of our Variance Exploding Stochastic Differential Equation (VE-SDE) diffusion model is applied, setting $(\sqrt{\lambda /\mu})^2=\sigma_t ^2$.
Furthermore, we can rewrite Eq. \eqref{eq:m1.prior_sub} as:
\begin{equation}
    \mathbf{z}_{t}=\mathbf{x}_{0|t} \approx \mathbf{x}_{t} + \sigma_t ^2 \mathbf{s}_{\boldsymbol{\theta} ^*}\left( \mathbf{x}_{t},t\right),
    \label{eq:m1.prior_sub17}
\end{equation}
{where $\sigma_t ^2$ is the noise level corresponding to timestep $t$ of the diffusion model.} This means that Eq. \eqref{eq:m1.prior_sub} has been transformed into a Tweedie denoising process in VE-SDE diffusion schema.
    {We then proceed from $\mathbf{z}_{t}$ (\textit{i.e.}, $\mathbf{x}_{0|t}$), to solve the data-fidelity subproblem shown in Eq. \eqref{eq:m1.data_sub} to a refined $\hat{\mathbf{x}}_{0|t}$ corrected with measurement $\mathbf{y}$. Note that we have $\mu =\lambda /\sigma _{t}^{2}$, then the data-fidelity subproblem can be rewritten to:
        \begin{equation}
            \hat{\mathbf{x}}_{0|t}=\underset{\mathbf{x}}{\mathrm{arg} \min}\left\| \mathbf{y}-\mathbf{A}\mathbf{x} \right\| ^2+\frac{\lambda}{2\sigma _{t}^{2}} \left\| \mathbf{x}-\mathbf{x}_{0|t} \right\| ^2,
            \label{eq:data_sub_final}
        \end{equation}
        where $\mu$ is a scaling factor hyper-parameter.}

However, due to the highly undercharacterized nature of the problem, it is difficult to produce the desired result in a single iteration. To take full advantage of the multi-step sampling of the diffusion model, we follow \cite{zhu2023denoising} to add noise to the refined $\hat{\mathbf{x}}_{0|t}$ with noise level corresponding to timestep $t$. The operation mentioned above can be succinctly formulated as follows:
\begin{equation}
    {\mathbf{x}}_{t-1}=\hat{\mathbf{x}}_{0\mid t}+\sigma_{t-1}\boldsymbol{\epsilon},
\end{equation}
which enables us to go back to the corresponding timestep of the diffusion model and continue sampling.

To illustrate our framework more clearly, we rewrite Eq. \eqref{eq:m1.hqs} as:
\begin{subequations}
    \label{eq:m1.hqs_final}
    \begin{numcases}{}
        \mathbf{x}_{0|t} = \mathbf{x}_{t} + \sigma_t ^2 \mathbf{s}_{\boldsymbol{\theta} ^*}\left( \mathbf{x}_{t},t\right)\label{eq:m1.prior_sub_final},\\
        \hat{\mathbf{x}}_{0|t}=\underset{\mathbf{x}}{\mathrm{arg} \min}\left\| \mathbf{y}-\mathbf{A}\mathbf{x} \right\| ^2+\frac{\lambda}{2\sigma _{t}^{2}} \left\| \mathbf{x}-\mathbf{x}_{0|t} \right\| ^2\label{eq:m1.data_sub_final},\\
        {\mathbf{x}}_{t-1}=\hat{\mathbf{x}}_{0\mid t}+\sigma_{t-1}\boldsymbol{\epsilon}\label{eq:m1.add_noise},
    \end{numcases}
\end{subequations}
where Eq. \eqref{eq:m1.prior_sub_final} denotes the prior subproblem, which estimates the noiseless image $\mathbf{x}_{0|t}$ via VE-SDE sampling with Tweedie's formula. Eq. \eqref{eq:m1.data_sub_final} is the data-fidelity subproblem, and Eq. \eqref{eq:m1.add_noise} is the forward sampling operation by adding noise at timestep $t$.
    {As mentioned earlier, the final convergence of the HQS-based penalty method to the optimal solution of the original problem presupposes that the penalty term $\mu$ gradually increases to $\infty$. In Eq. \eqref{eq:m1.hqs_final}, the penalty term is $\left(\lambda /2\sigma _{t}^{2} \right)$, where $\lambda$ is a fixed scaling factor, and $\sigma _{t}^{2}$ decreases progressively according to an elaborate schedule and eventually approaching zero. This leads to a gradual increase of the penalty term to infinity, thus allowing our method to gradually converge to the optimal solution. \textit{Notably, our design enables the adjustment of penalty term weights $\mu$ to be intrinsically linked to the diffusion model's sampling process. This approach eliminates the need for empirical adjustment of weights for different terms, thereby significantly enhancing the framework's utility and robustness.}}

For the data-fidelity subproblem as Eq. \eqref{eq:m1.data_sub_final}, we employ INR to address it. The data-fidelity term $\left\| \mathbf{y}-\mathbf{A}\mathbf{x} \right\| ^2$ ensures that the predicted $\mathbf{x}$ closely aligns with the measured data $\mathbf{y}$, thereby maintaining data fidelity. Additionally, the consistency term $\left\| \mathbf{x}-\mathbf{x}_{0|t} \right\| ^2$ emphasizes the similarity of the predicted image $\mathbf{x}$ to the prior image $\mathbf{x}_{0|t}$, which serves to reduce artifacts that might arise from overfitting to incomplete measurements $\mathbf{y}$.
A more detailed discussion of this topic can be found in Sec. \ref{subsec:INR}.

\par Fig. \ref{fig:method_overview} shows the overview of the proposed DPER. In our approach, the distribution prior sub-problem is solved by the diffusion model as shown in Fig. \ref{fig:method_overview}(a). The image prior $\mathbf{x}_T$ is initially sampled from a Gaussian distribution $\mathcal{N}(0, 1)$. Then, we generate a noisy prior image $\mathbf{x}_t$ at timestep $t$ by performing the reverse SDE process on a pre-trained score diffusion model. The generated noisy image $\mathbf{x}_t$ follows the pre-trained target distribution. Then we use Tweedie's formula~\cite{robbins1992empirical} to produce its denoised version $\mathbf{x}_{0\mid t}$, which serves as an initial solution to the distribution prior sub-problem in Eq. (\ref{eq:m1.prior_sub_final}).
As demonstrated in Fig. \ref{fig:method_overview}(b), the initialized prior image $\mathbf{x}_{0\mid t}$ is subsequently transferred to the INR refinement phase to address the data fidelity sub-problem in Eq. (\ref{eq:m1.data_sub_final}).
In the prior embedding step (Fig. \ref{fig:method_overview}(b.1)), INR is initially trained to present $\mathbf{x}_{0\mid t}$. Then it is further refined towards the measurement $\mathbf{y}$ to generate a high fidelity and realistic solution as in Fig. \ref{fig:method_overview}(b.2).
This refined image $\hat{\mathbf{x}}_{0\mid t}$ is the solution to the data fidelity sub-problem in Eq. (\ref{eq:m1.data_sub_final}).
$\hat{\mathbf{x}}_{0\mid t}$ is then transferred back to the SDE process to further enhance the target distribution prior.
By iteratively performing the reverse SDE sampling and INR refinement, the high-quality CT images can be reconstructed when the reverse SDE process is completed.

\subsection{Generative Image Prior by Diffusion Model}
\label{subsec:diffusion}
We employ a score-based diffusion model of the VE-SDE \cite{song2020score} form and utilize the reverse SDE sampling to address the distribution prior sub-problem in \eqnref{eq:m1.prior_sub_final}.
Specifically, the prior $\mathbf{x}_T$ is initialized with Gaussian distribution, and the denoising process based on score matching is executed sequentially with the timestep $T$ decay as follows:
\begin{equation}
    \mathbf{x}_{t-1}= \mathbf{x}_{t}+\left( \sigma _{t}^{2}-\sigma _{t-1}^{2} \right) \mathbf{s}_{\boldsymbol{\theta} ^*}\left( \mathbf{x}_{t},t \right)
    + \sqrt{\left( \sigma _{t}^{2}-\sigma _{t-1}^{2} \right)}\boldsymbol{\epsilon},
\end{equation}
where $\boldsymbol{\epsilon} \sim \mathcal{N}(\mathbf{0},\mathbf{I})$, $\sigma_t$ is the noise scale at timestep $t$ and $\mathbf{s}_{\boldsymbol{\theta} ^*}$ is the pre-trained score model.

When solving inverse problems via diffusion models, obtaining an estimated clean version of the sampled $\mathbf{x}_t$ before solving the data fidelity sub-problems has been shown to greatly improve performance \cite{chung2022improving, kawar2022denoising, chung2022diffusion, wang2022zero}.
Tweedie's formula \cite{robbins1992empirical} enables us to get the denoised result of $\mathbf{x}_t$ by computing the posterior expectation with one step, given a score function $\nabla _{\mathbf{x}_t}\log p\left(\mathbf{x}_t \right)$ as follows:
\begin{equation}
    \mathbb{E} \left[ \mathbf{x}_0\mid \mathbf{x}_t \right] = \mathbf{x}_t+\sigma_t^2 \nabla _{\mathbf{x}_t}\log p\left( \mathbf{x}_t \right) ,
\end{equation}
where $\sigma_t$ is the noise scale of $\mathbf{x}_t$.
The expression is the same as the prior subproblem Eq. \eqref{eq:m1.prior_sub_final} we derived in Sec. \ref{subsec:Overview}.
Given the estimated score function provided by the pre-trained score model $\mathbf{s}_{\boldsymbol{\theta}^*}$, we employ Tweedie's formula to get the noiseless prior image $\mathbf{x}_{0\mid t}$ from $\mathbf{x}_{t}$.
$\mathbf{x}_{0\mid t}$ denotes the denoised result inferred from $\mathbf{x}_t$ for clarity and conciseness.
The prior $\mathbf{x}_{0\mid t}$ is then employed to solve the data fidelity sub-problem with INR.

\subsection{Data Fidelity by Implicit Neural Representation}
\label{subsec:INR}
\par The data fidelity sub-problem in \eqnref{eq:m1.data_sub_final} is addressed by the unsupervised deep learning architecture INR. Specifically, as shown in Fig. \ref{fig:method_overview}(b.2), we model the underlying CT image $\mathbf{x}$ as an implicit continuous function that maps image coordinates to intensities. Then CT imaging forward model is performed to simulate the physical acquisition process from the image $\mathbf{x}$ to its according projection measurements. By minimizing the errors on the measurement domain, we can optimize an MLP network $\mathcal{F}_\mathbf{\Phi}$ to approximate the implicit function of $\mathbf{x}$.
The approximation of the implicit continuous function between coordinates and the reconstructed image enables a highly accurate simulation of the CT imaging forward model, thereby ensuring effective data fidelity with the incomplete measurement $\mathbf{y}$.

\par When addressing the challenging issue of high underdetermination caused by LACT and ultra-SVCT, we propose to guide INR towards a solution that maintains data fidelity and avoids degradation by incorporating the generative prior $\mathbf{x}_{0\mid t}$ sampled by the diffusion model.
Inspired by the INR work in~\cite{shen2022nerp}, we solve the data fidelity sub-problem in three stages: 1) Prior Embedding, 2) Data Fidelity Refinement, and 3) Image Reconstruction.

\subsubsection{Prior Embedding}
\par In this stage, our goal is to incorporate the generative prior $\mathbf{x}_{0\mid t}$ from the diffusion model into the INR model.
This strategy prevents INR from producing degradation reconstruction solution by avoiding overfitting to incomplete measurements $\mathbf{y}$. Technically, we represent the prior image $\mathbf{x}_{0\mid t}$ as a continuous function of image coordinate, which can be expressed as below:
\begin{equation}
    f: \mathbf{p}=(x,y)\in\mathbb{R}^2\rightarrow \mathbf{x}_{0\mid t}(\mathbf{p})\in\mathbb{R},
\end{equation}
where $\mathbf{p}$ is any spatial coordinate and $\mathbf{x}_{0\mid t}(\mathbf{p})$ is the corresponding intensity of the prior image at that position. As illustrated in Fig. \ref{fig:method_overview}(b.1), we leverage an MLP network $\mathcal{F}_\mathbf{\Phi}$ to approximate the function $f$ through optimizing the following objective function:
\begin{equation}
    \mathbf{\Phi}_{\mathrm{prior}}=\underset{\mathbf{\Phi }}{\mathrm{arg} \min} \frac{1}{N}\sum_{i=1}^N{\left\| \mathcal{F} _{\mathbf{\Phi }}\left( \mathbf{p}_{i} \right) -\mathbf{x}_{0\mid t}\left( \mathbf{p}_{i} \right) \right\| ^2},
    \label{prior_embedding}
\end{equation}
where $N$ denotes the number of sampling image coordinates at each iteration, $\mathbf{p}_{i}$ denotes the $i^{th}$ spatial coordinate. Once the optimization is completed, the prior image $\mathbf{x}_{0\mid t}(\mathbf{p})$ is embedded into the trainable weight $\mathbf{\Phi}_{\mathrm{prior}}$ of the MLP network.

From an initialization standpoint, we establish a viable starting point for the INR model, utilizing the generative prior offered by the diffusion model.

\subsubsection{Data Fidelity Refinement}
\par After the prior embedding, the INR network $\mathcal{F}_{\mathbf{\Phi}_{\mathrm{prior}}}$ incorporates the prior image $\mathbf{x}_{0\mid t}$. However, this prior image lacks data fidelity to the real incomplete measurement $\mathbf{y}$ since it is generated through unconditional random sampling. Therefore, we need to refine the MLP network $\mathcal{F}_{\mathbf{\Phi}_{\mathrm{prior}}}$ towards the measurement $\mathbf{y}$ by integrating the CT imaging model. Fig. \ref{fig:method_overview}(b.2) shows the refinement process. Technically, we sample coordinates $\mathbf{p}$ at a fixed interval $\Delta\mathbf{p}$ along an X-ray $\mathbf{r}$ and feed these coordinates into the MLP network to produce the corresponding intensities $I=\mathcal{F}_{\mathbf{\Phi}_{\mathrm{prior}}}(\mathbf{p})$. Then, we perform the CT imaging forward model (\textit{i.e.}, line integral transform) to generate the according measurement $\hat{\mathbf{y}}(\mathbf{r})$ for the X-ray $\mathbf{r}$. Finally, the MLP network $\mathcal{F}_{\mathbf{\Phi}_{\mathrm{prior}}}$ can be refined by minimizing the distance between the estimated measurement $\hat{\mathbf{y}}$ and real measurement $\mathbf{y}$. Formally, this process can be expressed as below:
\begin{equation}
    \begin{aligned}
        \mathbf{\Phi}_{\mathrm{refine}} & =\underset{\mathbf{\Phi}_{\mathrm{prior}}}{\mathrm{arg} \min}\frac{1}{|\mathcal{R} |}\sum_{\mathbf{r}\in\mathcal{R} } |\hat{\mathbf{y}}(\mathbf{r} )-\mathbf{y}(\mathbf{r} ) | \\
                                        & + \frac{\lambda}{2\sigma _{t}^{2}} \left\| \mathcal{F} _{\mathbf{\Phi }}\left( \mathbf{p}_{} \right) -\mathbf{x}_{0\mid t}\left( \mathbf{p}_{} \right) \right\| ^2,            \\
        \hat{\mathbf{y}}(\mathbf{r} )   & = \sum_{\mathbf{p}\in\mathbf{r}}\mathcal{F}_{\mathbf{\Phi}_\mathrm{prior}}(\mathbf{p})\cdot \Delta \mathbf{p} ,
    \end{aligned}
    \label{fidelity refine}
\end{equation}
where $\mathcal{R}$ denotes a set of sampling X-rays $\mathbf{r}$ at each epoch.
\subsubsection{Image Reconstruction}
\par Once the data fidelity refinement is completed, the MLP network $\mathcal{F}_{\mathbf{\Phi}_{\mathrm{refine}}}$ represents a solution $\hat{\mathbf{x}}_{0\mid t}$ that inherits the generative prior $\mathbf{x}_{0\mid t}$ from the diffusion model and is also highly consistent with the actual measurements $\mathbf{y}$.
Then, we reconstruct the solution $\hat{\mathbf{x}}_{0\mid t}(\mathbf{p})$ by feeding necessary image coordinates $\mathbf{p}$ into the MLP network $\mathcal{F}_{ \mathbf{\Phi}_{\mathrm{refine}}}$, \textit{i.e.}, $\hat{\mathbf{x}}_{0\mid t}=\mathcal{F}_{\mathbf{\Phi}_\mathrm{refine}}(\mathbf{p})$.

\begin{algorithm}[!t]
    \caption{DPER}
    \begin{algorithmic}[1]
        \Require Measurements $\mathbf{y}$, Pre-trained score function $\mathbf{s}_{\boldsymbol{\theta}^*}$, MLP network $\mathcal{F} _{\mathbf{\Phi }}$,  Timesteps $T$, Refinement interval $\texttt{RI}$, Noise schedule $\{\sigma_t\}_{t=0}^{T}$, Coordinates of target images $\mathbf{p}$, Sampled X-rays set $\mathcal{R}$.
        \State $\mathbf{x}_T \sim ({\bf 0}, \sigma_{T}^{2}\mathbf{I})$;
        \For{$t = T,\dots,1$} \do \\
        \LineComment{\textcolor[rgb]{0.40,0.40,0.40}{\textit{1. Distribution Prior Sub-problem}}}
        \State $\boldsymbol{\epsilon}\thicksim \mathcal{N} \,\,(\mathbf{0},\mathbf{I})$
        \State {$\mathbf{x}_{t-1}= \mathbf{x}_{t}+\left( \sigma _{t}^{2}-\sigma _{t-1}^{2} \right) \mathbf{s}_{\boldsymbol{\theta} ^*}\left( \mathbf{x}_{t},t \right)
            + \sqrt{\left( \sigma _{t}^{2}-\sigma _{t-1}^{2} \right)}\boldsymbol{\epsilon}$}
        \If{$(t-1)\ \mathrm{mod}\ \texttt{RI} =0$}
        \State {$\mathbf{x}_{0\mid t}=\mathbf{x}_t+\sigma _{t}^{2}\mathbf{s}_{\boldsymbol{\theta} ^*}\left( \mathbf{x}_t,t \right)$}
        \Comment{\textcolor[rgb]{0.40,0.40,0.40}{\textit{Tweedie's formula}}}
        \LineComment{\textcolor[rgb]{0.40,0.40,0.40}{\textit{2. Data Fidelity Sub-problem}}}
        \State {$\mathbf{\Phi}_{\mathrm{prior}}=\underset{\mathbf{\Phi }}{\mathrm{arg} \min} \frac{1}{N}\sum_{i=1}^N{\left\| \mathcal{F} _{\mathbf{\Phi }}\left( \mathbf{p} \right) -\mathbf{x}_{0\mid t}\left( \mathbf{p} \right) \right\| ^2}$}
        \State {{$\mathbf{\Phi}_{\mathrm{refine}}= \underset{\mathbf{\Phi}_{\mathrm{prior}}}{\mathrm{arg} \min}\frac{1}{|\mathcal{R} |}\sum_{\mathbf{r}\in\mathcal{R} } |\hat{\mathbf{y}}(\mathbf{r} )-\mathbf{y}(\mathbf{r} ) | + \frac{\lambda}{2\sigma _{t}^{2}} \left\| \mathcal{F} _{\mathbf{\Phi }}\left( \mathbf{p}_{} \right) -\mathbf{x}_{0\mid t}\left( \mathbf{p}_{} \right) \right\| ^2$} }
        \State {$\hat{\mathbf{x}}_{0\mid t}= \mathcal{F}_{\mathbf{\Phi}_\mathrm{refine}}(\mathbf{p})$}
        \State {${\mathbf{x}}_{t-1}=\hat{\mathbf{x}}_{0\mid t}+\sigma _{t-1}\boldsymbol{\epsilon}$, {$\boldsymbol{\epsilon}\thicksim \mathcal{N} \,\,(\mathbf{0},\mathbf{I})$ }}
        \Comment{\textcolor[rgb]{0.40,0.40,0.40}{\textit{Add Noise}}}
        \EndIf
        \EndFor
        \State \textbf{return} $\mathbf{x}_0$
    \end{algorithmic}\label{alg:DPER}
\end{algorithm}

\subsection{Alternative Optimization Strategy}
\par Following the HQS algorithm, we iteratively optimize the diffusion model-based regularization sub-problem and the INR-based data fidelity sub-problem using an alternative approach.
As shown in Algorithm \ref{alg:DPER}, we intersperse the solution of the data fidelity sub-problem with the reverse SDE sampling process (as distribution prior sub-problem).
A hyper-parameter refinement interval $\texttt{RI}\in\mathbb{R}$ is introduced for balancing the weights of the two sub-problems in the overall process. $\texttt{RI}$ enables the data fidelity sub-problem to be executed at a specified time step.
When $(t-1)\mod \texttt{RI}\ne 0$, our algorithm solely performs reverse SDE denoising to generate images that match the target distribution, \textit{i.e.}, solving the distribution prior sub-problem.
When $(t-1)\mod \texttt{RI}=0$, we solve the data fidelity sub-problem once at the current timestep. Specifically, an estimated clean image at $t$ is obtained using \eqnref{eq:m1.prior_sub_final}.
Then, INR is applied to optimize the data fidelity sub-problem given the measurement $\mathbf{y}$ and prior image ${\mathbf{x}}_{0\mid t}$, as explained in Sec. \ref{subsec:INR}.
Finally, the corresponding level of noise is added to the result $\hat{\mathbf{x}}_{0\mid t}$ to return to the manifold at timestep $t-1$ and continue back to the reverse SDE sampling. We can trade-off between reconstruction performance and computational consumption by tuning the $\texttt{RI}$ as discussed in Sec. \ref{subsec:RI}.

\subsection{Implementation Details}
\par For the Diffusion Sampling stage (Fig. \ref{fig:method_overview}(a)), we set the total number of the sampling $T$ as 2000 and the refinement interval $\texttt{RI}$ as 50. We adopt the \verb|ncsnpp|~\cite{song2020score} network as the score-based diffusion model architecture without any modification.
During the prior embedding and data fidelity refinement, the numbers of training iterations are 50 and 250, the batch sizes are 1000 (\textit{i.e.}, $N=1000$  in Eq. \eqref{prior_embedding}) and 128 (\textit{i.e.}, $|\mathcal{R}|=128$ in Eq. \eqref{fidelity refine}), and the learning rates are 1e-3 and 1e-4. Adam optimizer with default hyper-parameters is leveraged. \textit{Note that all hyper-parameters are tuned on two samples of the AAPM dataset \cite{mccollough2017low} and are then held constant across all other samples.}

\section{Experiments}
\label{sec:experiments}
\begin{figure*}[!t]
	\centerline{\includegraphics[width=1\textwidth]{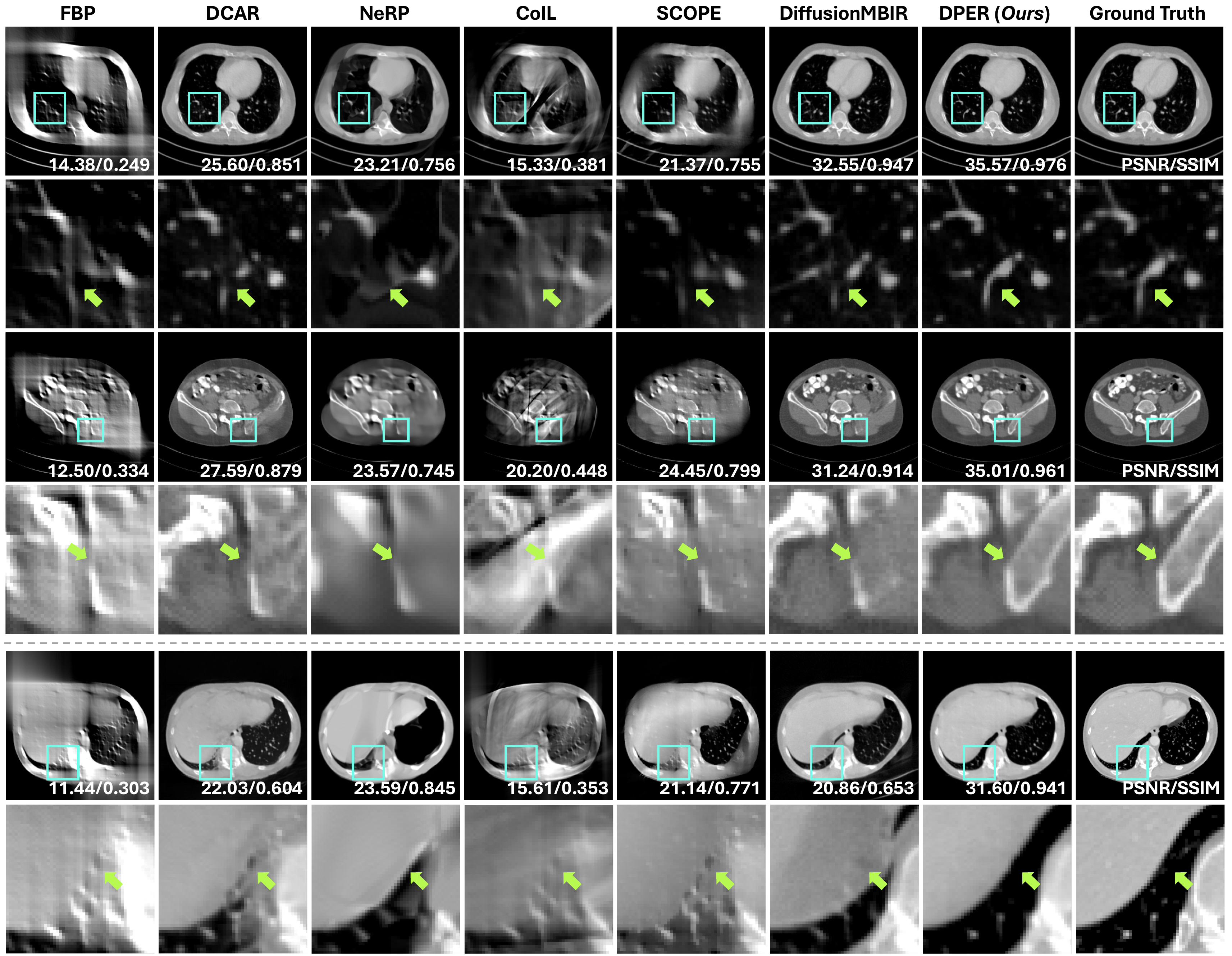}}
	\caption{{Qualitative results of methods in comparison on two test samples (\#0 (Rows 1-2), \#239 (Rows 3-4) of the AAPM dataset and \#1 (Rows 5-6)) of the LIDC dataset for the LACT task of [0, 90]° scanning range.}}
	\label{fig:Exp_LACT}
\end{figure*}

\begin{table*}
	\centering
	\caption{Quantitative results of methods in comparison on the AAPM and LIDC datasets for the LACT tasks of [0, 90]°, [0, 120]°, and [0, 150]° scanning ranges. The best and second best results are highlighted in \textbf{bold} and \underline{underline}, respectively.}
	\label{tab:LACT}
	\begin{tabular}{cccccccc}
		\toprule
		\multirow{2}{*}{\textbf{Dataset}} & \multirow{2}{*}{\textbf{Method}}     & \multicolumn{2}{c}{[0, 90]°} & \multicolumn{2}{c}{[0, 120]°} & \multicolumn{2}{c}{[0, 150]°}                                                                                              \\
		\cmidrule{3-8}
		                                  &                                      & PSNR$\uparrow$               & SSIM$\uparrow$                & PSNR$\uparrow$                & SSIM$\uparrow$                & PSNR$\uparrow$             & SSIM$\uparrow$                \\
		\midrule
		\multirow{8}{*}{\textbf{AAPM}}    & FBP\cite{fbp}                                  & 13.73$\pm$0.52               & 0.3015$\pm$0.0419             & 16.97$\pm$0.63                & 0.3995$\pm$0.0350             & 22.41$\pm$1.25             & 0.5002$\pm$0.0353             \\
		                                  & DCAR\cite{huang2019data}             & 28.27$\pm$1.39               & 0.8796$\pm$0.0180             & 34.36$\pm$1.60                & 0.9426$\pm$0.0108             & 38.69$\pm$1.05             & 0.9658$\pm$0.0086             \\
		                                  & DDNet\cite{zhang2018sparse}          & 26.85$\pm$1.43               & 0.7342$\pm$0.0300             & 31.00$\pm$1.54                & 0.8259$\pm$0.0265             & 34.92$\pm$1.50             & 0.8812$\pm$0.0211             \\
		                                  & NeRP\cite{shen2022nerp}              & 24.34$\pm$2.05               & 0.7511$\pm$0.0626             & 31.96$\pm$3.19                & 0.9077$\pm$0.0354             & 36.98$\pm$3.07             & 0.9441$\pm$0.0260             \\
		                                  & CoIL\cite{sun2021coil}               & 18.44$\pm$1.62               & 0.4002$\pm$0.0521             & 20.99$\pm$1.85                & 0.5398$\pm$0.0409             & 28.40$\pm$1.95             & 0.7936$\pm$0.0484             \\
		                                  & SCOPE\cite{wu2022self}               & 23.77$\pm$1.98               & 0.8079$\pm$0.0306             & 30.45$\pm$2.72                & 0.9312$\pm$0.0161             & 37.67$\pm$1.46             & 0.9752$\pm$0.0039 \\
		                                  & MCG\cite{chung2022improving}         & 32.66$\pm$2.35               & 0.9306$\pm$0.0120             & 34.93$\pm$1.41                & 0.9445$\pm$0.0065             & 35.78$\pm$0.78             & 0.9486$\pm$0.0063             \\
		                                  & DiffusionMBIR\cite{chung2023solving} & \underline{35.07$\pm$2.26}   & \underline{0.9483$\pm$0.0084} & \underline{37.21$\pm$2.92}    & \underline{0.9660$\pm$0.0057} & \underline{42.34$\pm$0.82} & \underline{0.9802$\pm$0.0063}             \\
		                                  & DPER (Ours)                          & \textbf{37.20$\pm$2.38}      & \textbf{0.9759$\pm$0.0064}    & \textbf{43.17$\pm$1.91}       & \textbf{0.9902$\pm$0.0027}    & \textbf{50.01$\pm$1.30}    & \textbf{0.9978$\pm$0.0015}    \\
		\midrule
		\multirow{8}{*}{\textbf{LIDC}}    & FBP\cite{fbp}                                  & 12.79$\pm$0.94               & 0.2753$\pm$0.0273             & 15.71$\pm$1.02                & 0.4508$\pm$0.0338             & 19.02$\pm$1.23             & 0.5614$\pm$0.0465             \\
		                                  & DCAR\cite{huang2019data}             & 20.84$\pm$1.50               & 0.5909$\pm$0.0487             & 24.29$\pm$1.78                & 0.6867$\pm$0.0522             & 27.76$\pm$1.87             & 0.7465$\pm$0.0536             \\
		                                  & DDNet\cite{zhang2018sparse}          & 20.20$\pm$1.65               & 0.5951$\pm$0.0425             & 24.31$\pm$1.48                & 0.7066$\pm$0.0464             & 27.66$\pm$1.31             & 0.7419$\pm$0.0527             \\
		                                  & NeRP\cite{shen2022nerp}              & 22.53$\pm$1.00               & 0.7163$\pm$0.0194             & 26.37$\pm$1.89                & 0.8128$\pm$0.0303             & 31.32$\pm$1.14             & 0.8723$\pm$0.0168             \\
		                                  & CoIL\cite{sun2021coil}               & 15.93$\pm$0.59               & 0.3619$\pm$0.0493             & 17.03$\pm$0.96                & 0.4520$\pm$0.0438             & 22.76$\pm$1.55             & 0.6142$\pm$0.0389             \\
		                                  & SCOPE\cite{wu2022self}               & 19.67$\pm$1.13               & 0.6456$\pm$0.0358             & 26.92$\pm$1.58                & 0.8785$\pm$0.0307             & \underline{38.36$\pm$1.75} & \underline{0.9818$\pm$0.0073} \\
		                                  & MCG\cite{chung2022improving}         & 21.58$\pm$1.01               & 0.7910$\pm$0.0189             & 24.64$\pm$2.51                & 0.8623$\pm$0.0341             & 32.52$\pm$2.03             & 0.9296$\pm$0.0177             \\
		                                  & DiffusionMBIR\cite{chung2023solving} & \underline{23.27$\pm$1.47}   & \underline{0.8197$\pm$0.0278} & \underline{28.93$\pm$1.68}    & \underline{0.9170$\pm$0.0215} & 35.67$\pm$1.69             & 0.9548$\pm$0.0125             \\
		                                  & DPER (Ours)                          & \textbf{28.96$\pm$1.29}      & \textbf{0.9134$\pm$0.0223}    & \textbf{34.76$\pm$1.72}       & \textbf{0.9655$\pm$0.0092}    & \textbf{42.42$\pm$1.86}    & \textbf{0.9882$\pm$0.0041}    \\
		\bottomrule
	\end{tabular}
\end{table*}

\subsection{Experimental Settings}
\subsubsection{Dataset and Pre-Processing}
\paragraph{Simulated Dataset} We conduct simulation experiments on two well-known public CT datasets, including AAPM 2016 low-dose CT grand challenge \cite{mccollough2017low} and Lung Image Database Consortium (LIDC) image collection dataset \cite{armato2011lung}. The AAPM dataset DICOM Full Dose data was acquired at 120kV and 200mAs in the portal venous phase using a Siemens SOMATOM Flash scanner.
The LIDC database is not performed specifically for the purpose of the database so a heterogeneous range of scanner models (\textit{e.g.}, GE Medical Systems LightSpeed scanner models, Toshiba Aquilion scanners, Philips Brilliance scanner models) and technical parameters (\textit{e.g.}, 120-140kV tube peak potential energies, 40-627 mA tube current range, 0.6mm-5.0mm slice thickness) were represented. There are evident differences in the acquisition protocol and subjects between LIDC and AAPM datasets. Therefore, the evaluation based on the LIDC dataset can effectively reflect the generalization and robustness of the reconstruction methods in OOD settings. For the AAPM dataset, we extract 5376 2D slices from CT volumes of 9 subjects and randomly split them into 4240 training slices and 1136 validation slices. We then use 101 slices from an unseen volume as test data. For the LIDC dataset, we randomly choose 30 slices from various subjects as additional test data. Moreover, we retrospectively employ radon transform implemented by \texttt{scikit-image} library of Python (\textit{i.e}., parallel X-ray beam) to generate the incomplete (SV and LA) projection data. \textit{Note that the training and validation images from only the AAPM dataset are used for training three supervised baselines (FBPConvNet, DCAR, and DDNet)}.

\paragraph{Clinical Dataset with Lesion} We also conduct experimental validation on an in-house COVID-19 chest clinical CT dataset. All subjects were patients with severe COVID-19 who underwent serial chest CT scans during hospital stay. The examinations were performed with an Optima CT680 CT scanner (GE Medical Systems, USA;). The CT protocol was as follows: the tube voltage was 120 kVp, X-ray tube current was automatic (180$\sim$400 mA). The distance between the source and the detector was 94.9 cm, and the distance between the source and patient was 54.1 cm, respectively. The image matrix featured dimensions of 512$\times$512 pixels. The LACT and SVCT projections were extracted from the rebind fan-beam full-view projection data.
To validate the robustness and performance of our method, we seamlessly applied the score function trained on the AAPM dataset to clinical data.
	{\paragraph{Real Projection Dataset} In order to more closely match the real clinical scenario settings, we also conduct an experiment on the real projections data from the {\texttt{LDCT-and-Projection-data}}\footnote{https://www.cancerimagingarchive.net/collection/ldct-and-projection-data/}~\cite{moen2021low}. It contains de-identified CT projection data and acquisition geometry collected in helical mode from either a GE Discovery CT750i, SOMATOM Definition AS+ or SOMATOM Definition Flash CT system. The tube potential is 80$\sim$120 kV, and the field-of-view (FOV) diameter covers an expansive 282$\sim$500 mm. Using the rebin technique described in \cite{noo1999single}, we obtain fan-beam projections with 2304 uniformly distributed scanning views over a complete 360-degree range. As with the COVID-19 dataset, the model trained on simulated data is seamlessly applied to the real projection data.}

\subsubsection{Methods in comparison}
\par Eight representative CT reconstruction approaches from three categories are utilized as comparative methods: 1) one model-based method (FBP \cite{fbp}); 2) three supervised CNN-based DL methods (FBPConvNet \cite{jin2017deep}, DDNet \cite{zhang2018sparse}, and DCAR \cite{huang2019data}); 3) Five unsupervised DL methods (SCOPE \cite{wu2022self}, NeRP \cite{shen2022nerp}, CoIL \cite{sun2021coil}, MCG \cite{chung2022improving} and DiffusionMBIR \cite{chung2023solving}). Among them, FBPConvNet is solely utilized for the SVCT task, while DCAR is exclusively applied to the LACT task. We implement the three supervised methods according to the original papers and subsequently train them on the training and validation sets of the AAPM dataset using the Adam optimizer \cite{kingma2014adam} with a mini-batch size of 16 and a learning rate of 1e-4. The training epochs for FBPConvNet and DCAR are set to 50, while for DDNet, it is set to 500. It is important to emphasize that these hyperparameters have been meticulously fine-tuned on the AAPM dataset to ensure a fair comparison. As for the five fully unsupervised methods (SCOPE, NeRP, CoIL, MCG, and DiffusionMBIR) are evaluated using their official code. As diffusion-based models, MCG, DiffusionMBIR, and our DPER use the same score function checkpoint provided by Chung et al. \cite{chung2023solving}, which is pre-trained on the AAPM dataset. For the three fully unsupervised INR-based methods, we carefully tune their hyper-parameters on two samples from the AAPM datasets and then fix these parameters for all other cases.

\begin{figure*}[!t]
	\centerline{\includegraphics[width=1\textwidth]{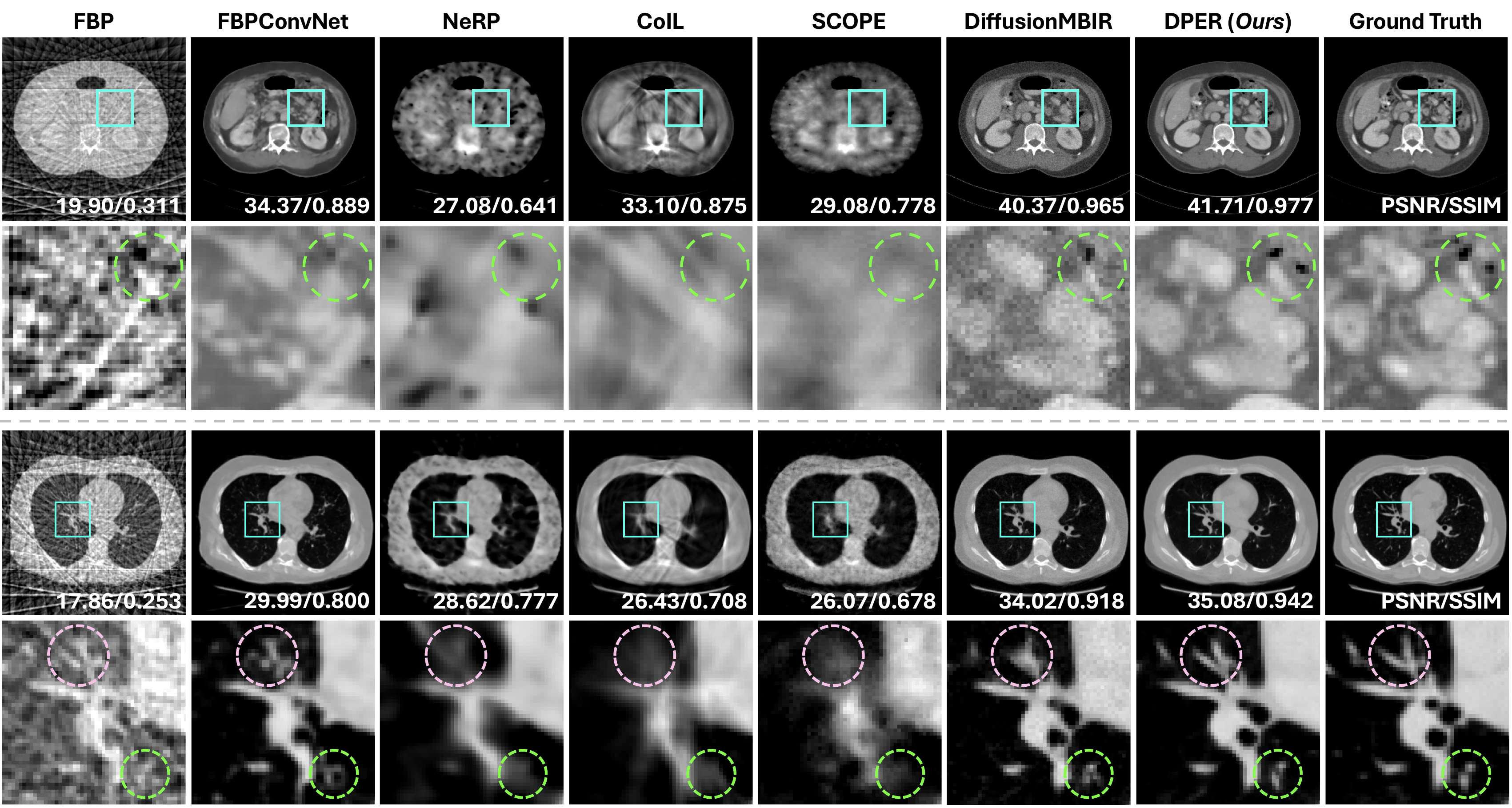}}
	\caption{
		{Qualitative results of methods in comparison on two test samples (\#133 (Rows 1-2) and \#1 (Rows 3-4)) of the AAPM and LIDC datasets for the ultra-SVCT task of 20 input views.}}
	\label{fig:Exp_SVCT}
\end{figure*}

\begin{table*}
\centering
\caption{Quantitative results of methods in comparison on the AAPM and LIDC datasets for the ultra-SVCT tasks of 10, 20, and 30 input views. The best and second best results are highlighted in \textbf{bold} and \underline{underline}, respectively.}
\label{tab:SVCT}
\begin{tabular}{cccccccc} 
\toprule
\multirow{2}{*}{\textbf{Dataset}} & \multirow{2}{*}{\textbf{Method}} & \multicolumn{2}{c}{\textbf{10 views}}                                                                                                     & \multicolumn{2}{c}{\textbf{20 views}}                                                    & \multicolumn{2}{c}{\textbf{30 views}}                                                     \\ 
\cmidrule{3-8}
                                  &                                  & PSNR$\uparrow$                            & SSIM$\uparrow$                                                                                & PSNR$\uparrow$                            & SSIM$\uparrow$                               & PSNR$\uparrow$                            & SSIM$\uparrow$                                \\ 
\midrule
\multirow{8}{*}{\textbf{AAPM}}    & FBP~\cite{fbp}                              & 13.54$\pm$0.71                            & 0.1725$\pm$0.0082                                                                             & 18.78$\pm$0.86                            & 0.2993$\pm$0.0162                            & 22.63$\pm$0.89                            & 0.4107$\pm$0.0158                             \\
                                  & FBPConvNet~\cite{jin2017deep}                       & 28.56$\pm$1.01                            & 0.7950$\pm$0.0292                                                                             & 32.84$\pm$1.19                            & 0.8751$\pm$0.0225                            & 35.29$\pm$1.33                            & 0.9127$\pm$0.0201                             \\
                                  & DDNet\cite{zhang2018sparse}                            & 26.54$\pm$0.91                            & 0.7005$\pm$0.0277                                                                             & 30.43$\pm$0.97                            & 0.7905$\pm$0.0214                            & 32.82$\pm$1.29                            & 0.8516$\pm$0.0232                             \\
                                  & NeRP\cite{shen2022nerp}                             & 19.63$\pm$2.35                            & 0.3864$\pm$0.0843                                                                             & 25.67$\pm$1.52                            & 0.6139$\pm$0.0597                            & 29.98$\pm$1.47                            & 0.7694$\pm$0.0456                             \\
                                  & CoIL\cite{sun2021coil}                             & 25.46$\pm$1.00                            & 0.6841$\pm$0.0363                                                                             & 31.62$\pm$1.32                            & 0.8563$\pm$0.0220                            & 34.74$\pm$1.32                            & 0.9099$\pm$0.0147                             \\
                                  & SCOPE\cite{wu2022self}                            & 22.13$\pm$0.67                            & 0.5525$\pm$0.0287                                                                             & 28.10$\pm$0.70                            & 0.8249$\pm$0.0135                            & 32.22$\pm$1.19                            & 0.8596$\pm$0.0190                             \\
                                  & MCG\cite{chung2022improving}                              & 34.27$\pm$0.65                            & 0.9202$\pm$0.0094                                                                             & 35.22$\pm$0.57                            & 0.9327$\pm$0.0084                            & 35.62$\pm$0.56                            & 0.9384$\pm$0.0086                             \\
                                  & DiffusionMBIR\cite{chung2023solving}                    & \underline{37.20$\pm$1.26}                    & \underline{0.9531$\pm$0.0065}                                                                     & \underline{39.88$\pm$1.03}                    & \underline{0.9651$\pm$0.0054}                    & \underline{40.78$\pm$0.80}                    & \underline{0.9694$\pm$0.0044}                     \\
                                  & {DPER (Ours)}    & {\textbf{38.16$\pm$0.82}} & \textbf{{0.9663$\pm$0.0054}} & {\textbf{41.06$\pm$0.85}} & {\textbf{0.9771$\pm$0.0043}} & {\textbf{42.32$\pm$0.94}} & {\textbf{0.9846$\pm$0.0035}}  \\ 
\midrule
\multirow{8}{*}{\textbf{LIDC}}    & FBP~\cite{fbp}                              & {13.76$\pm$1.11}          & {0.1609$\pm$0.0314}                                                           & {18.31$\pm$1.01}          & {0.2854$\pm$0.0384}          & {20.96$\pm$0.83}          & {0.3465$\pm$0.0167}           \\
                                  & FBPConvNet~\cite{jin2017deep}                       & {25.64$\pm$0.80}          & {0.6972$\pm$0.0244}                                                           & {30.91$\pm$1.11}          & {0.8124$\pm$0.0211}          & {34.05$\pm$1.09}          & {0.8925$\pm$0.0148}           \\
                                  & DDNet\cite{zhang2018sparse}                            & {24.62$\pm$0.87}          & {0.7012$\pm$0.0239}                                                           & {28.81$\pm$0.92}          & {0.8185$\pm$0.0178}          & {29.44$\pm$0.50}          & {0.8593$\pm$0.0124}           \\
                                  & NeRP\cite{shen2022nerp}                             & {20.92$\pm$1.34}          & {0.5635$\pm$0.0569}                                                           & {27.33$\pm$1.36}          & {0.7481$\pm$0.0421}          & {31.74$\pm$2.13}          & {0.8502$\pm$0.0474}           \\
                                  & CoIL\cite{sun2021coil}                             & {25.63$\pm$1.83}          & {0.8141$\pm$0.0344}                                                           & {29.79$\pm$1.18}          & {0.8222$\pm$0.0381}          & {32.19$\pm$1.08}          & {0.8775$\pm$0.0262}           \\
                                  & SCOPE\cite{wu2022self}                            & {23.69$\pm$1.20}          & {0.7052$\pm$0.0432}                                                           & {26.49$\pm$0.97}          & {0.7682$\pm$0.0158}          & {28.34$\pm$1.15}          & {0.8662$\pm$0.0374}           \\
                                  & MCG\cite{chung2022improving}                              & {30.33$\pm$1.01}          & {0.8526$\pm$0.0187}                                                           & {33.81$\pm$0.99}          & {0.8862$\pm$0.0211}          & {35.48$\pm$1.14}          & {0.9091$\pm$0.0144}           \\
                                  & DiffusionMBIR\cite{chung2023solving}                    & \underline{{31.49$\pm$1.69}}  & \underline{{0.8730$\pm$0.0090}}                                                   & \underline{{34.67$\pm$1.07}}  & \underline{{0.9069$\pm$0.0122}}  & \underline{{36.60$\pm$1.12}}  & \underline{{0.9396$\pm$0.0104}}   \\
                                  & DPER (Ours)                      & \textbf{{32.52$\pm$1.11}} & \textbf{{0.9142$\pm$0.0133}}                                                  & \textbf{{36.10$\pm$1.18}} & \textbf{{0.9441$\pm$0.0047}} & \textbf{{37.79$\pm$1.02}} & \textbf{{0.9566$\pm$0.0137}}  \\
\bottomrule
\end{tabular}
\end{table*}

\subsubsection{Evaluation Metrics}
\par We utilize two standard metrics to quantitatively assess the performance of the compared methods: Peak Signal-to-Noise Ratio (PSNR) and Structural Similarity Index Measure (SSIM).

\subsection{Comparison with SOTA Methods for LACT Task}
\par We compare the proposed DPER model with the eight baselines on the AAPM and LIDC datasets for the LACT reconstructions with scanning angles of  [0, 90]°, [0, 120]°, and [0, 150]° scanning ranges. Table \ref{tab:LACT} shows the quantitative results. On the AAPM dataset, our DPER, DiffusionMBIR and MCG produce the top-three performances and are significantly superior to the other baselines, benefiting from the generative prior provided by the diffusion model. However, on the LIDC dataset, we observe that DiffusionMBIR and MCG, trained on the AAPM dataset, experience severe performance degradation due to the OOD problem. In some cases, their results were even worse than naive INR-based SCOPE, for example, 35.67 dB compared to 38.36 dB in PSNR for the [0, 150]° range. In contrast, the performance of our method remains stable over two different test datasets.
Compared with DiffusionMBIR and MCG, DPER achieves superior results for the demanding OOD LIDC dataset by utilizing the INR method, which guarantees dependable data fidelity with the CT measurement data while maintaining the prior local image consistency.
\par Fig. \ref{fig:Exp_LACT} demonstrates the qualitative results on two representative samples (\#0 and \#1) from the AAPM and LIDC datasets. Upon visual inspection, it is evident that the model-based FBP and three INR-based methods (NePR, CoIL, and SCOPE) fail to yield satisfactory results on either dataset. This limitation is primarily attributed to the significant data gap inherent in the LACT problem. Although DCAR and DiffusionMBIR produce satisfactory results on the AAPM dataset, their reconstructions on the LIDC dataset exhibit noticeable distortions, primarily attributed to the OOD problem. Conversely, the CT images generated by our DPER model across both datasets demonstrate high image quality in terms of both local details and global structural integrity.

\subsection{Comparison with SOTA Methods for Ultra-SVCT Task}
\par We conduct a comparative evaluation of our DPER model against eight baseline methods on the AAPM and LIDC datasets for ultra-SVCT reconstructions with 10, 20, and 30 input views. We present the quantitative results in Table \ref{tab:SVCT}. The three INR-based methods hardly produce pleasing results for the ultra-SVCT task. Supervised FBPConvNet and DDNet perform well on the AAPM dataset while suffering from severe performance drops on the OOD LIDC dataset. DPER, DiffusionMBIR and MCG all obtain excellent and robust results on two datasets.

Fig. \ref{fig:Exp_SVCT} shows the qualitative result on two representative samples (\#133 and \#1) of AAPM and LIDC datasets. The results from the FBP algorithm suffer severe streaking artifacts, while INR-based methods (NeRP, CoIL, and SCOPE) produce overly smooth artifacts results.
INR-based methods tend to overfit the incomplete measurement and lack data distribution prior information, resulting in less detailed tissue reconstruction compared to the supervised method FBPConvNet. FBPConvNet yields relatively acceptable results. However, the structure precision remains somewhat unclear. In the ultra-SVCT task, DPER and DiffusionMBIR obtain the best and second-best performance. {Compared to DiffusionMBIR, our DPER method is superior in both quantitative metrics and qualitative visualization, exhibiting less noise and more accurate structures.}
The findings from LIDC show that the three diffusion-based methods produce favorable reconstruction results, even when dealing with OOD data in the ultra-SVCT task.

\subsection{Comparison on Clinical COVID-19 Dataset}

\begin{figure}[!t]
	\centerline{\includegraphics[width=0.5\textwidth]{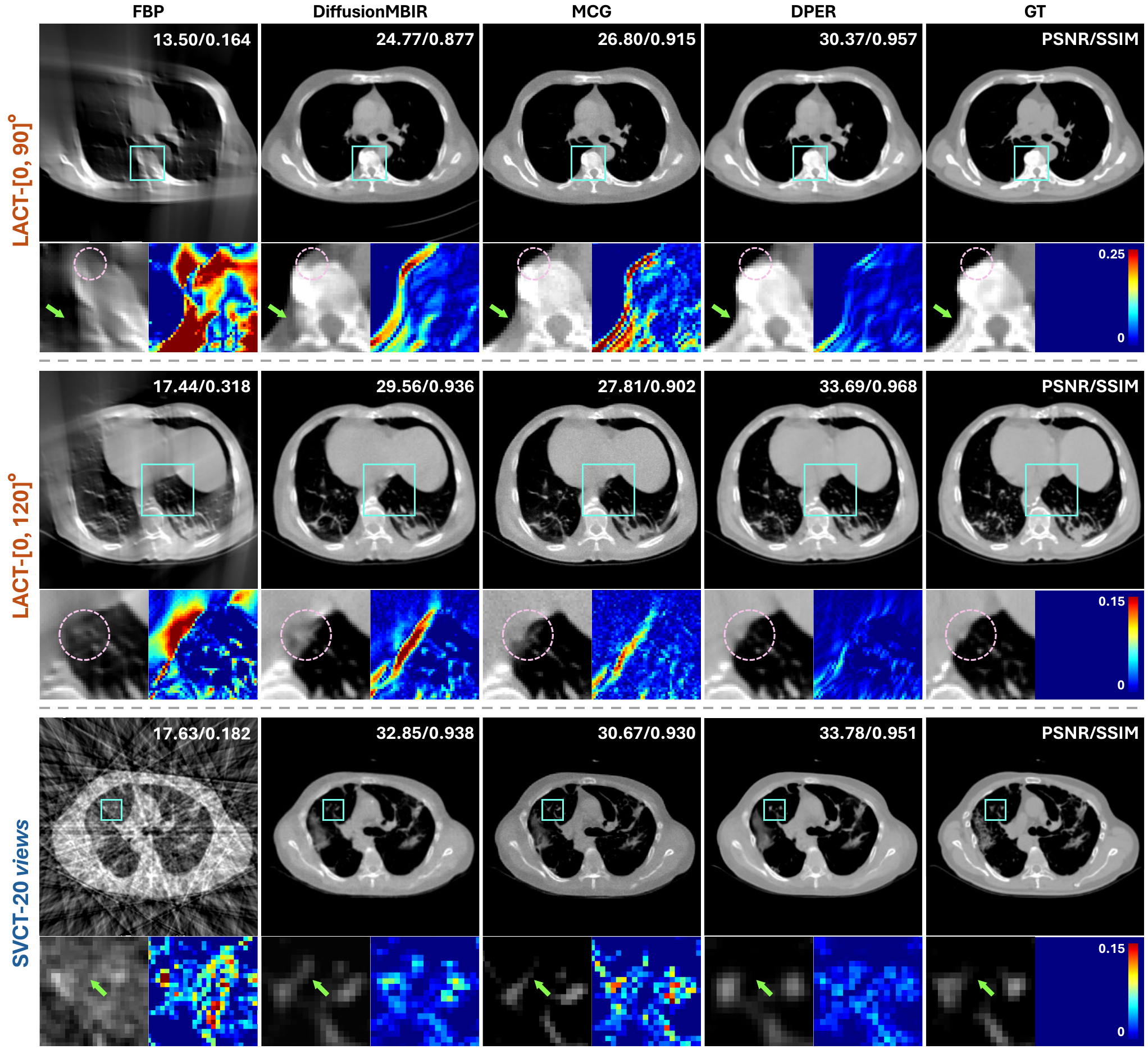}}
	\caption{{Comparison of LACT task ([0, 90]° and [0, 120]°) and SVCT task (20 views) on test samples (\#90, \#178 and \#213) of the clinical COVID-19 CT dataset.}}
	\label{fig:Exp-COVID-19}
\end{figure}

\begin{table}
\centering

\caption{Quantitative results (PSNR / SSIM) of different methods on the clinical COVID-19 dataset and real LDCT-PD projection dataset for LACT task (90° and 120°) and SVCT task (20 views and 30 views). The Best performances are highlighted in \textbf{bold}.}
\label{tab: real_data}
\resizebox{\columnwidth}{!}{

\begin{tabular}{cccccc} 
\toprule
\multirow{2}{*}{\textbf{Dataset}}  & \multirow{2}{*}{\textbf{Method}} & \multicolumn{2}{c}{\textbf{LACT}}                                   & \multicolumn{2}{c}{\textbf{SVCT}}                           \\ 
\cmidrule(l){3-6}
                                   &                                  & {[}0, 90]°                       & {[}0, 120]°                      & 20 \textit{views}                & 30 \textit{views}        \\ 
\midrule
\multirow{4}{*}{\textbf{COVID-19}} & FBP~\cite{fbp}                              & 12.01 / 0.2370                   & 14.48 / 0.2788                   & 11.51 / 0.1492                   & 21.85 / 0.2700           \\
                                   & MCG~\cite{chung2022improving}                              & 27.59 / 0.9153                   & 29.08 / 0.9207                   & ~30.37 / 0.9082                  & 32.14 / 0.9228           \\
                                   & DiffusionMBIR~\cite{chung2023solving}                    & 26.73 / 0.8879                   & 29.54 / 0.9292                   & 30.96 / 0.9119                   & 32.93 / 0.9314           \\
                                   & DPER (\textit{Ours})             & \textbf{31.63} / \textbf{0.9548} & \textbf{33.07} / \textbf{0.9678} & \textbf{31.92} / \textbf{0.9286} & \textbf{34.75 / 0.9466}  \\ 
\midrule
\multirow{4}{*}{\textbf{LDCT-PD}}  & FBP~\cite{fbp}                              & 13.08 / 0.2814                   & 16.09 / 0.3467                   & 16.19 / 0.0979                   & 18.55 / 0.1438           \\
                                   & MCG~\cite{chung2022improving}                              & 23.61 / 0.8515                   & 27.86 / 0.8916                   & 29.08 / 0.8972                   & 31.10 / 0.9038           \\
                                   & DiffusionMBIR~\cite{chung2023solving}                    & 21.46 / 0.8482                   & 25.52 / 0.8947                   & 29.75 / 0.8937                   & 31.60 / 0.9097           \\
                                   & DPER (\textit{Ours})             & \textbf{28.21 / 0.9139}          & \textbf{32.60 / 0.9439}          & \textbf{30.84 / 0.9123}          & \textbf{32.80 / 0.9226}  \\
\bottomrule
\end{tabular}

}

\end{table}

To conduct a more comprehensive assessment of the effectiveness of our method, we carried out experimental validation using the clinical COVID-19 CT dataset. We compared the proposed DPER with the current SOTA method {MCG} and DiffusionMBIR on LACT task (with 90° and 120° scanning, respectively) and SVCT task (with 20 views and {30 views projection, respectively}). We present the quantitative results in Table \ref{tab: real_data}.

Fig.~\ref{fig:Exp-COVID-19} shows the qualitative results on some samples (\#90, \#178 and \#213) of the clinical COVID-19 CT dataset.
Confronted with this challenging OOD scenario (different acquisition protocols, anomalous structures caused by diseases), the proposed DPER consistently demonstrates commendable generalization performance. As depicted in the first row in Fig. \ref{fig:Exp-COVID-19}, for the 90° LACT task, our DPER significantly outperforms the SOTA methods DiffusionMBIR and MCG, both in quantitative metrics and visual representation. Specifically, in areas highlighted by pink circles and green arrows, the reconstruction results from DiffusionMBIR either omit or distort the original bone structure. In contrast, the results obtained from DPER closely align with the ground truth. This is further corroborated by the 120° LACT task results, as illustrated in Fig. \ref{fig:Exp-COVID-19}.
{In cases presenting anomalous structures associated with COVID-19, our DPER method exhibited superior capability in reconstructing the original content of these structures with high fidelity.}
In contrast, the DiffusionMBIR method exhibited blurring and inaccuracies in reconstructing these anomalous areas. In summary, DPER can still hold outstanding performance in the more complex OOD clinical dataset without any fine-tuning on pre-trained diffusion prior.

\subsection{Comparison on Real Projection Dataset}
\begin{figure}[!t]
	\centerline{\includegraphics[width=0.5\textwidth]{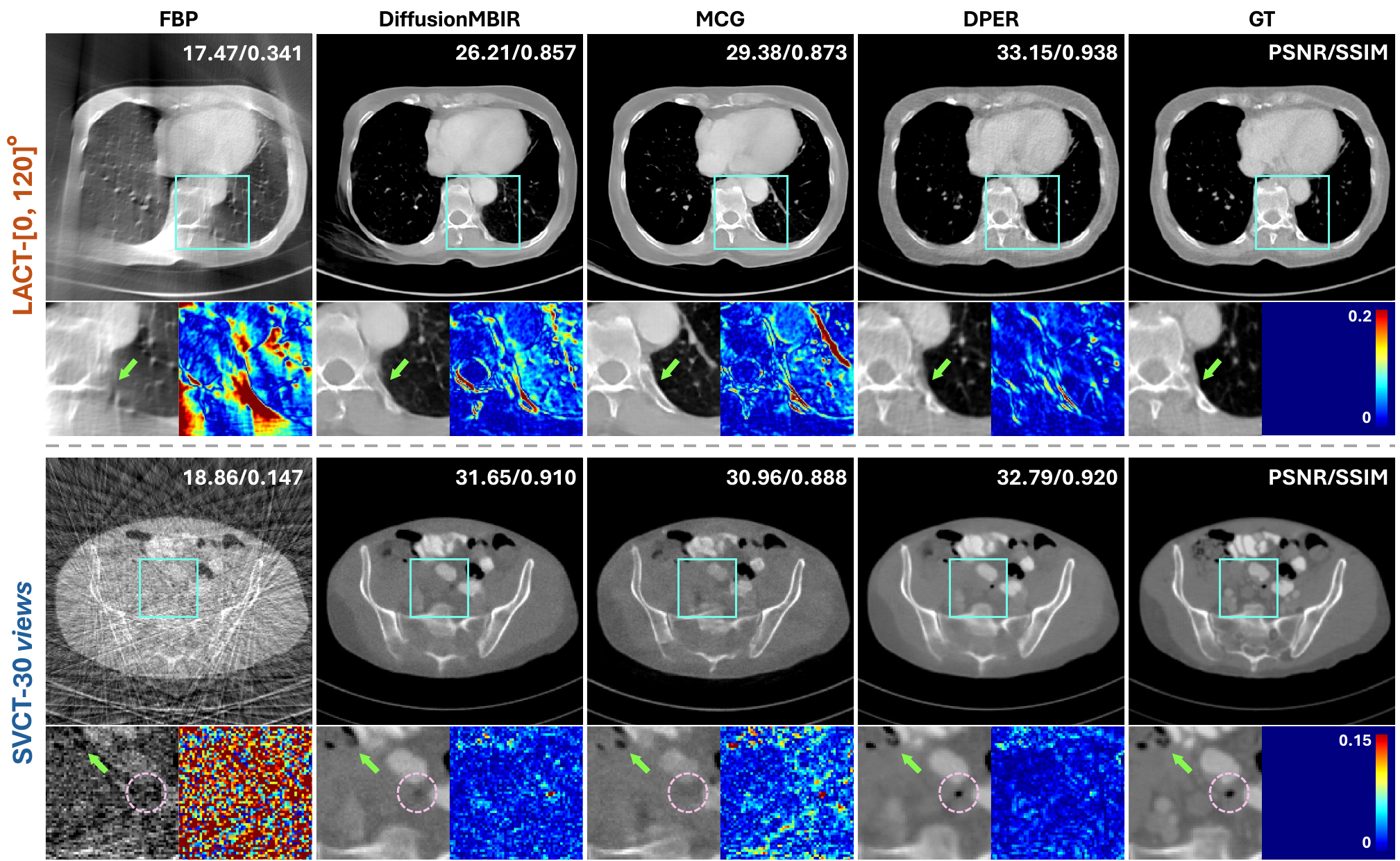}}
	\caption{{Comparison of LACT task ([0, 120]°) and SVCT task (30 views) on representative samples of the real projection clinical LDCT-PD dataset.}}
	\label{fig:Exp-LDCT-PD}
\end{figure}
The proposed DPER is also evaluated on the real projections data from the \texttt{LDCT-and-Projection-data} dataset to more closely match challenging real-world clinical scenarios. Raw projection data makes the reconstruction task challenging due to non-eliminable noise in the acquisition process, complex acquisition protocols and even possible geometric errors. We compared the proposed DPER with two SOTA methods, MCG and DiffusionMBIR, on the same settings as the COVID-19 dataset.

Quantitative and qualitative results are displayed in Table \ref{tab: real_data}. and Fig. \ref{fig:Exp-LDCT-PD}, respectively. It is clear that the FBP results for both the LACT and SVCT tasks have difficulty in partitioning structure and detail due to severe artifacts.
For the 120° LACT task, scrutinizing the specific ROIs in Fig. \ref{fig:Exp-LDCT-PD}, we find that DiffusionMBIR is unable to restore the correct shape of the bone. Additionally, its reconstruction results show a severe defect in the lower left. This may be due to its departure from the prior manifold of the data, caused by the continuous forced application of data consistency operations. MCG obtains slightly better results with Manifold Constraints~\cite{chung2022improving}, but demonstrates poor data fidelity. In particular, the green arrows mark the wrong bone and muscle structures. In contrast, the proposed DPER is highly consistent with GT both in terms of prior and measurement manifold. Notably, a close inspection of the reconstruction results reveals that both DiffusionMBIR and MCG exhibit cartoon-like features that blur fine structures. This blurring is due to the disparity between the pre-trained score function and the target distribution of the reference image. The proposed DPER, on the other hand, can fully utilize the respective strengths of the data prior and measurement data to produce accurate reconstructions with clear details. The 30-views SVCT task shows similar findings. In the regions indicated by the pink circles and green arrows, both DiffusionMBIR and MCG introduced blurred or erroneous structures as well as significant noise, while DPER was closer to the reference image.

\subsection{Effectiveness of Refinement Interval}
\label{subsec:RI}
\begin{figure}[!t]
	\centerline{\includegraphics[width=0.5\textwidth]{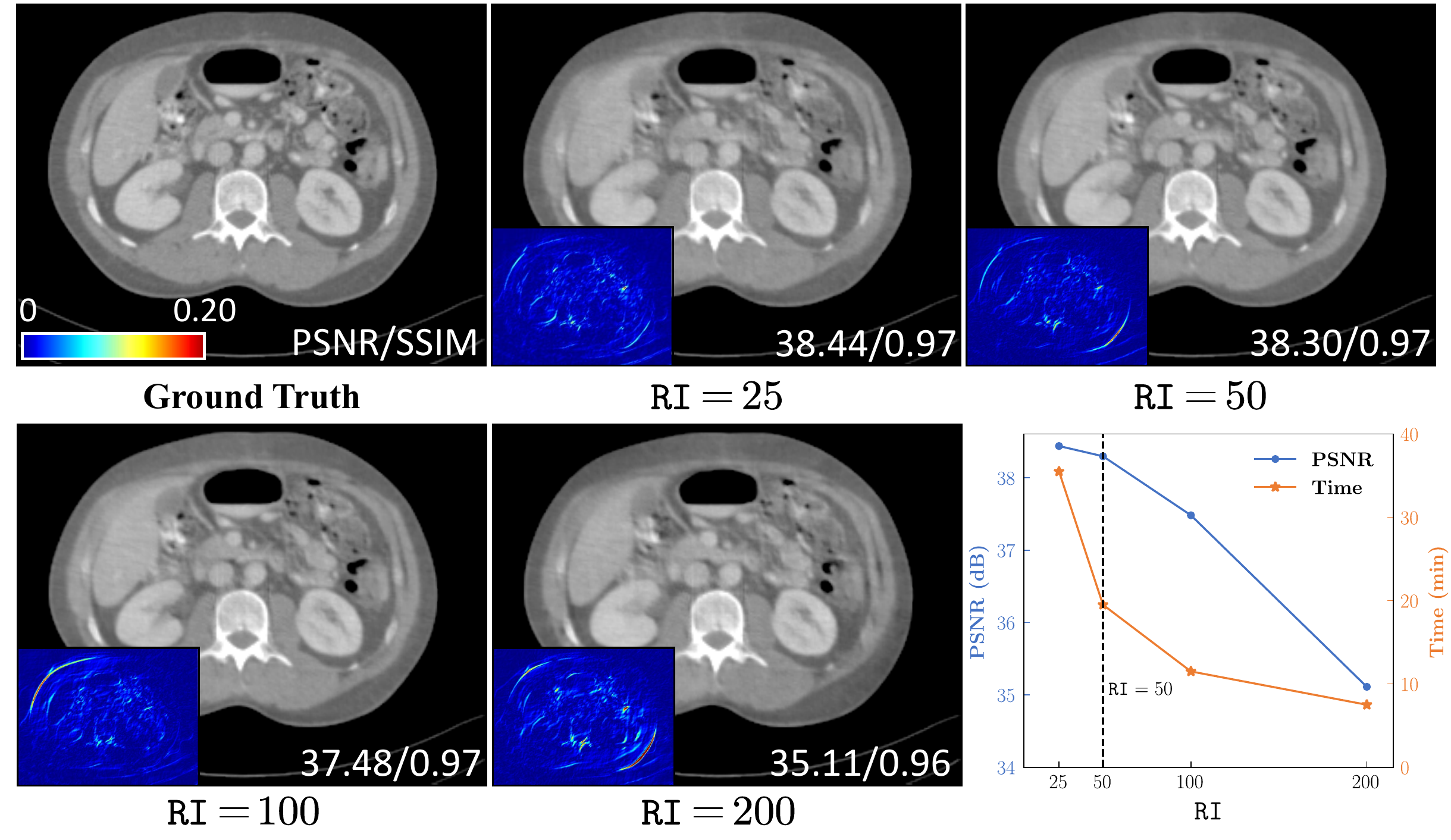}}
	\caption{Qualitative and quantitative results of the DPER with five different refinement intervals $\texttt{RI}$ on a test sample (\#133) of the AAPM dataset for the LACT task of [0, 90]° scanning range.}
	\label{fig:Exp_ABL_RI_all}
\end{figure}
\par In the proposed framework, the hyperparameter refinement interval \texttt{RI} balances the contribution of the data fidelity sub-problem and the distribution prior sub-problem in Eq. \eqref{eq:m1.hqs_final}. To evaluate its impact, we set four distinct \texttt{RI} values (25, 50, 100, and 200) on the AAPM dataset for the LACT task with the scanning range of [0, 90]°. Fig. \ref{fig:Exp_ABL_RI_all} shows the model performance and time cost curve varying with the refinement interval and the qualitative comparisons.
It is clear that as the refinement interval increases, the performance of the model gradually decreases, but the running time of the model increases considerably.
As the refinement interval increases, the influence of the distribution prior regularization becomes more pronounced in the solution of the overall inverse problem, while the impact of the data fidelity subproblem diminishes.
	{Consequently, the reconstructed image may deviate from the ground truth.}
The visualization results are consistent with the quantitative analysis.
In short, refinement intervals provide us with a balanced choice between performance and time consumption.
It is worth noting that when the $\texttt{RI}\leqslant 50$, further reduction of the refinement interval results in a very limited performance gain (\textit{e.g.}, only 0.14 improvement in PSNR) but a multiplicative computational cost. Therefore, in this paper we set $\texttt{RI}=50$.

\subsection{Comparison of MBIR and INR as Data-fidelity Solver}
\label{subsec:Diffusion Generative Prior}
\begin{figure}[!t]
	\centerline{\includegraphics[width=0.5\textwidth]{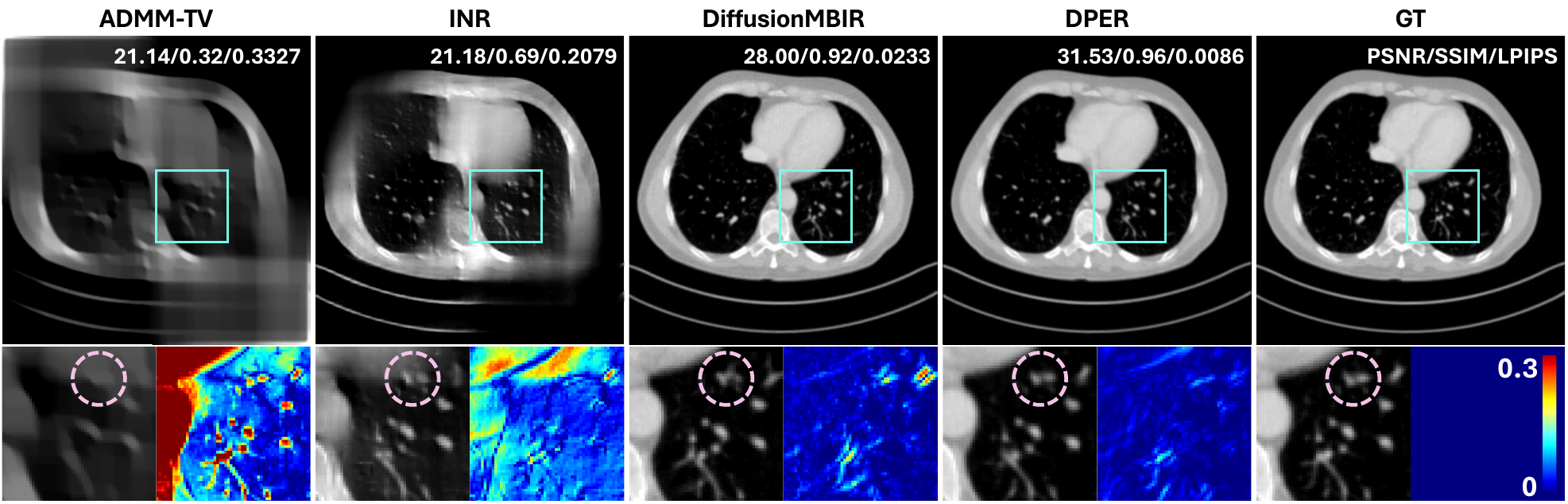}}
	\caption{{Qualitative and quantitative results of MBIR (ADMM-TV) and INR with and without score-based prior (DiffusionMBIR is ADMM-TV with score-based prior, and DPER is INR with score-based prior) on one test sample(\# 8)) of the AAPM dataset for the LACT task of [0, 90]° scanning range.}}
	\label{fig:Exp_ABL_wo_diffusion}
\end{figure}

\par To explore the advantages of the proposed DPER in solving the data-fidelity subproblem as Eq. \eqref{eq:m1.data_sub_final}, we evaluate the performance of traditional MBIR and INR-based methods under two distinct scenarios: one involving the application of a diffusion model as a prior augmentation, and the other without it. Specifically, ADMM-TV and SCOPE have been selected as prototypical examples of MBIR and INR-based methods, respectively. Furthermore, DiffusionMBIR and proposed DPER are employed as representatives of the two methods in the case with diffusion prior. Fig. \ref{fig:Exp_ABL_wo_diffusion} displays example results for the LACT task with a scanning range of [0, 90]°. In the absence of a diffusion prior, the MBIR method tends to exhibit vast divergence phenomenon in tissue edges or boundaries, rendering fine structures challenging to discern. Conversely, owing to the inherent continuity prior, INR-based methods demonstrate a marked efficacy in constraining the solution space for the data-fidelity subproblem. This is particularly noticeable in regions that are not directly visible and in the delineation of fine structures. With the introduction of the diffusion prior, this advantage is directly reflected in a better solution for the data-fidelity subproblem (\textit{i.e.}, Eq. \eqref{eq:m1.data_sub_final}) in the overall framework as Eq. \eqref{eq:m1.hqs_final}.
Additionally, the reconstruction results and error map show that the INR provides a higher data fidelity than ADMM-TV, which leads the DPER to hold a better authenticity than DiffusionMBIR in the detailed area.

\subsection{Effectiveness of Tweedie Denoising}
\begin{figure}[!t]
	\centerline{\includegraphics[width=0.5\textwidth]{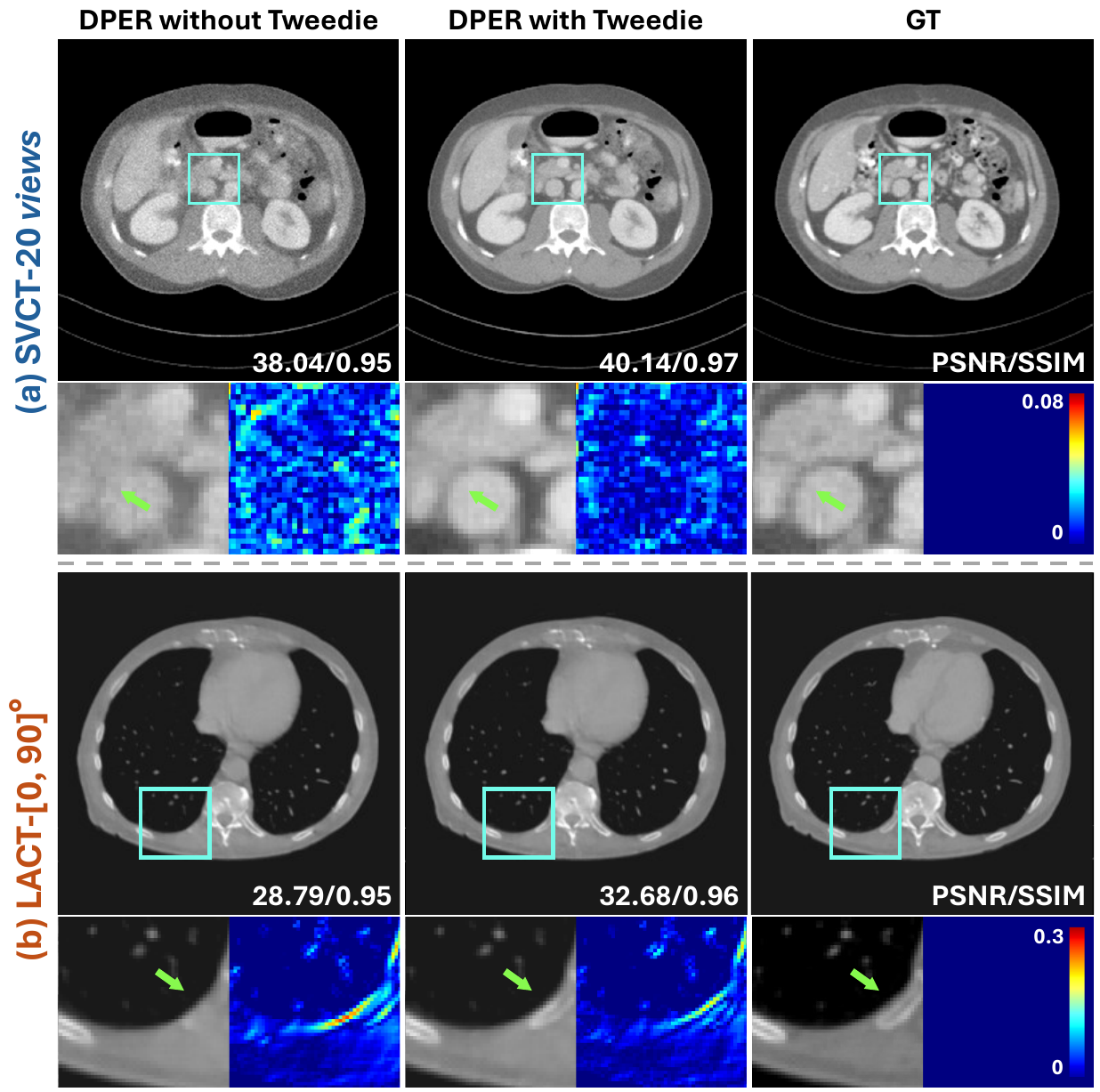}}
	\caption{Comparison of LACT task of [0, 90]° and SVCT task of 20 views with and without Tweedie denoising in the proposed DPER.}
	\label{fig:Exp_ABL_Tweedie}
\end{figure}

\par  To fully illustrate the benefits of Tweedie's strategy for $\mathbf{x}_{0}$ estimation, we conducted ablation experiments to compare the difference between using $\mathbf{x}_{0}$ (\textit{i.e.}, $\left\| \mathbf{A}\hat{\mathbf{x}}_{0|t}-\mathbf{y} \right\| _{2}^{2}$) and $\mathbf{x}_{t}$ (\textit{i.e.}, $\left\| \mathbf{A}{\mathbf{x}}_{t}-\mathbf{y} \right\| _{2}^{2}$ for solving the data-fidelity subproblem. Fig. \ref{fig:Exp_ABL_Tweedie} displays the qualitative results of these two schemes on the AAPM dataset of the LACT task of [0, 90]° scanning range and SVCT task of 20 views.
In the SVCT task, the reconstruction results of DPER without Tweedie denoising exhibit significant blur (as shown in the zoomed-in area of Fig. \ref{fig:Exp_ABL_Tweedie}(a)). These results are characterized by indistinct boundaries between tissues and the presence of noticeable noise, which compromises the clarity and distinctness of the reconstructed images. In contrast, the reconstruction of DPER with Tweedie denoising is much clearer. The 90° LACT task shows a similar phenomenon; for example, in Fig. \ref{fig:Exp_ABL_Tweedie}(b), the blue arrow indicates that Tweedie denoising provides more precise information.
The DPER that incorporates Tweedie demonstrates superior performance over the DPER without Tweedie for both the LACT and SVCT tasks.
Recently, a lot of work has been demonstrating that using Tweedie denoising helps improve the performance of solving inverse problems via diffusion model~\cite{chung2022improving,wang2022zero}. Indeed, using estimated $\mathbf{x}_{0}$ for data consistency operations (\textit{i.e.}, $\left\| \mathbf{A}\hat{\mathbf{x}}_{0|t}-\mathbf{y} \right\| _{2}^{2}$)) has become a widely used and recognized paradigm in the field of solving inverse problems via diffusion model \cite{chung2022improving,wang2022zero,chung2022diffusion,zhu2023denoising}.

\subsection{Effectiveness of Alternating Optimization Strategy}
\begin{figure}[!t]
	\centerline{\includegraphics[width=0.5\textwidth]{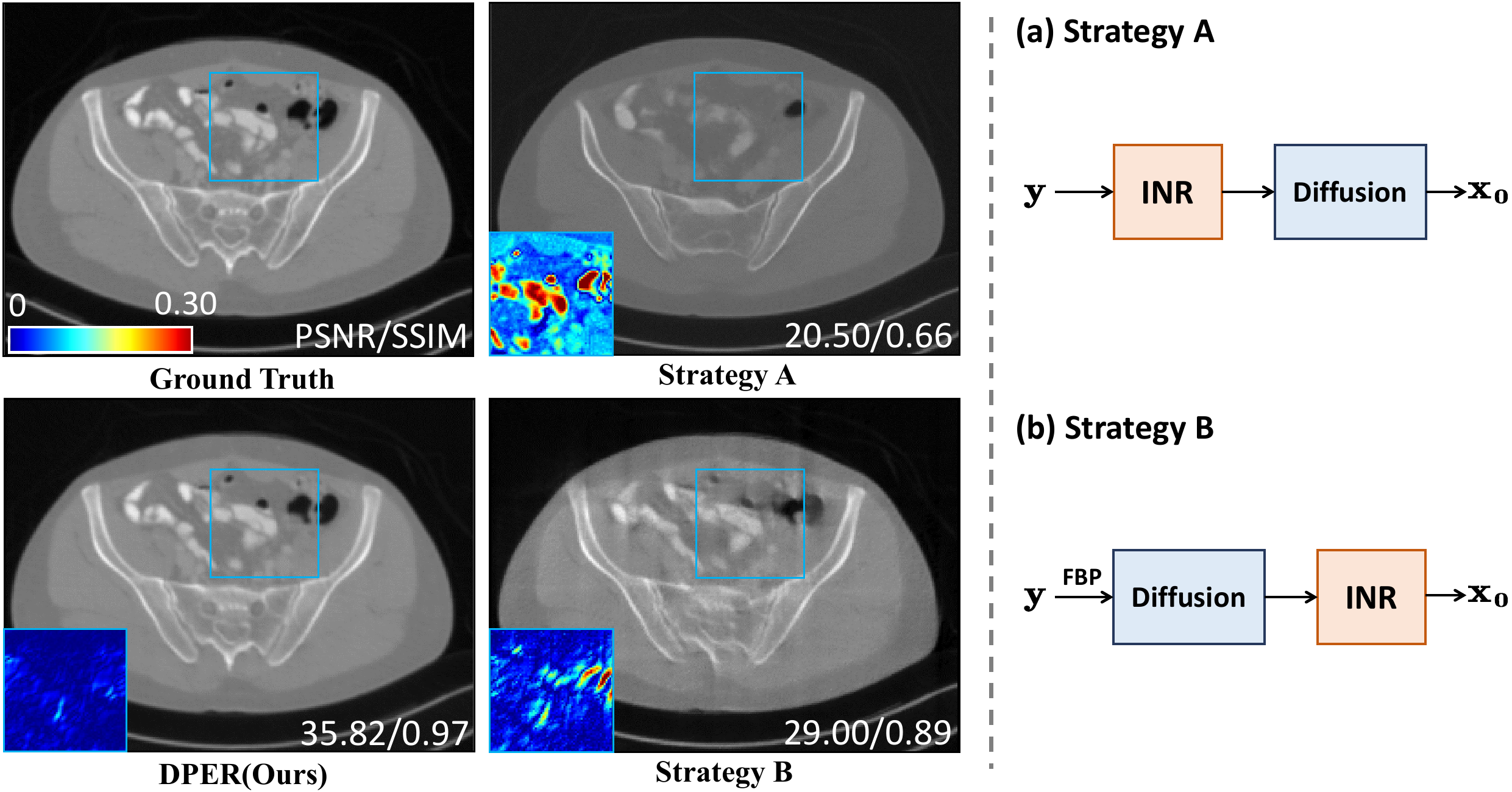}}
	\caption{Qualitative results of different strategies on a test sample (\#276) of the AAPM dataset for the LACT task of [0, 90]° scanning range. The right side shows the flowchart of strategies A and B.}
	\label{fig:Exp_ABL_Strategy_all}
\end{figure}

\par To evaluate the effectiveness of alternative optimization, we compared two naive strategies (as shown in Fig.~\ref{fig:Exp_ABL_Strategy_all}) that directly integrate the INR and diffusion model~\cite{muller2023diffrf, gu2023nerfdiff} without alternative optimization of the proposed framework.
	{ Strategy A: INR is first parameterized by directly fitting the incomplete measurement data. Then, the pre-trained diffusion model performs conditional sampling to refine the image provided by INR, producing the reconstruction result.
		Strategy B: First, the pre-trained diffusion model performs conditional sampling based on the FBP result of the incomplete measurement. Then, the generated result is refined by INR, followed by the Implicit Neural Representation Refinement process of DPER, to produce the reconstruction result. The conditional sampling process in Strategy B is conducted by adding an appropriate level of noise to the input FBP image, followed by reverse diffusion sampling.}

Fig. \ref{fig:Exp_ABL_Strategy_all} shows the performance of the proposed alternate optimization outperforms strategies A and B.
From the error map, the reconstructed result of strategy A is greatly different from ground truth, which can be caused by insufficient data fidelity constraints in generating.
In contrast, strategy B achieves improved data fidelity but retains some artifacts.
DPER with an alternative optimization strategy yields accurate and high-fidelity reconstructed results.
This suggests that the alternative optimization effectively balances the prior and data fidelity terms, leading to high-fidelity results in CT reconstruction.

\subsection{Evaluating Robustness of the Reconstruction}

\subsubsection{The uncertainty of reconstruction}
\begin{figure}[!t]
	\centerline{\includegraphics[width=0.5\textwidth]{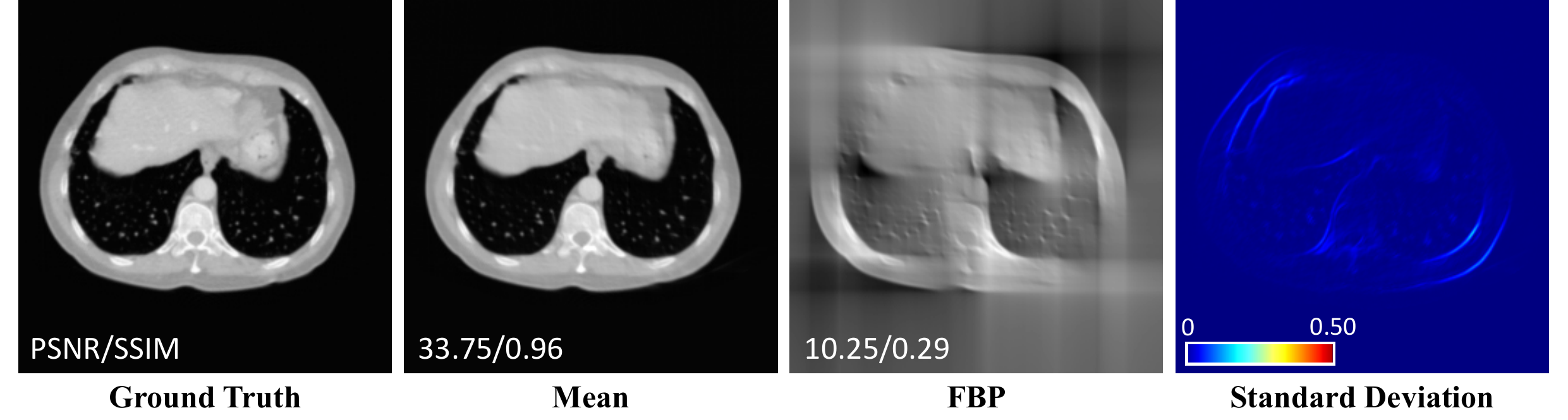}}
	\caption{Qualitative results of quantifying uncertainty of reconstruction on a test sample (\#40) of the AAPM dataset for the LACT task of [0, 90°] scanning range. From left to right: Ground Truth, mean, FBP result and standard deviation with ten reconstructions.}
	\label{fig:result_abl_uncertainty}
\end{figure}

Due to the generative nature of diffusion models, DPER can produce multiple feasible reconstructed results with slightly different results based on the same incomplete measurement.
To evaluate the robustness and qualify the uncertainty of reconstructions, we perform ten reconstructions simultaneously on the same incomplete measurements and calculate the mean and standard deviation of the reconstructions.
In Fig. \ref{fig:result_abl_uncertainty}, the mean of reconstructions is close to ground truth.
In the standard deviation map, the portion with actual measurements has a lower value, which proves that the INR refinement provides efficient data fidelity constraint in reconstruction.
In conclusion, DPER can reduce the uncertainty of reconstruction, which can provide reliable results.

\subsubsection{Influence of Noises}
During the process of CT acquisition, the presence of noise is an inevitable phenomenon, attributable to a myriad of factors.
	{We conducted noise interference experiments by introducing noise to the projection data to verify the robustness of the proposed method. Specifically, we performed experiments on the LACT task with a 90° range and the SVCT task with 20 views, introducing Poisson noise (modeling photon variability noise) and Gaussian noise (modeling electronic system readout noise), respectively.}
For Poisson noise, we accomplished this through the use of a statistical Poisson model, as delineated below:
\begin{equation}
	\mathbf{Y}(\theta,\rho) \sim \mathrm{Poisson} \left\{b \times e^{\mathbf{y}(\theta,\rho)} + r \right\},
	\label{poisson model}
\end{equation}
$\mathbf{Y(\theta,\rho)}$ represents the transmitted X-ray photon intensity, where $b$ denotes the incident X-ray photon intensity, and $r$ symbolizes the average of the background events and read-out noise variance. To simulate different Signal-to-Noise Ratio (SNR) levels, we assign the value of $r$ as 10, while $b$ is set at $1.3 \times 10^{4}$, $4 \times 10^{4}$, and $4 \times 10^{5}$, corresponding to approximate SNR levels of 32dB, 37 dB, and 43dB, respectively. The resulting noisy sinograms, denoted as $\mathbf{y}_{\theta}'$, are computed using the negative logarithm transformation: $\mathbf{y}_{\theta}' = -\ln(\mathbf{Y}(\theta,\rho)/b)$.
	{For Gaussian noise, we followed the setting in~\cite{zhang2024wavelet}, adding Gaussian noise with zero mean and 0.64 variance to the projected data.}
In our study, we evaluate the performance of our proposed DPER method against three baseline methodologies: FBP, SCOPE (as the SOTA in INR-based methods), and DiffusionMBIR (as the SOTA in diffusion-based methods) for CT image recovery from these noisy sinograms.

\begin{figure}[!t]
	\centerline{\includegraphics[width=0.5\textwidth]{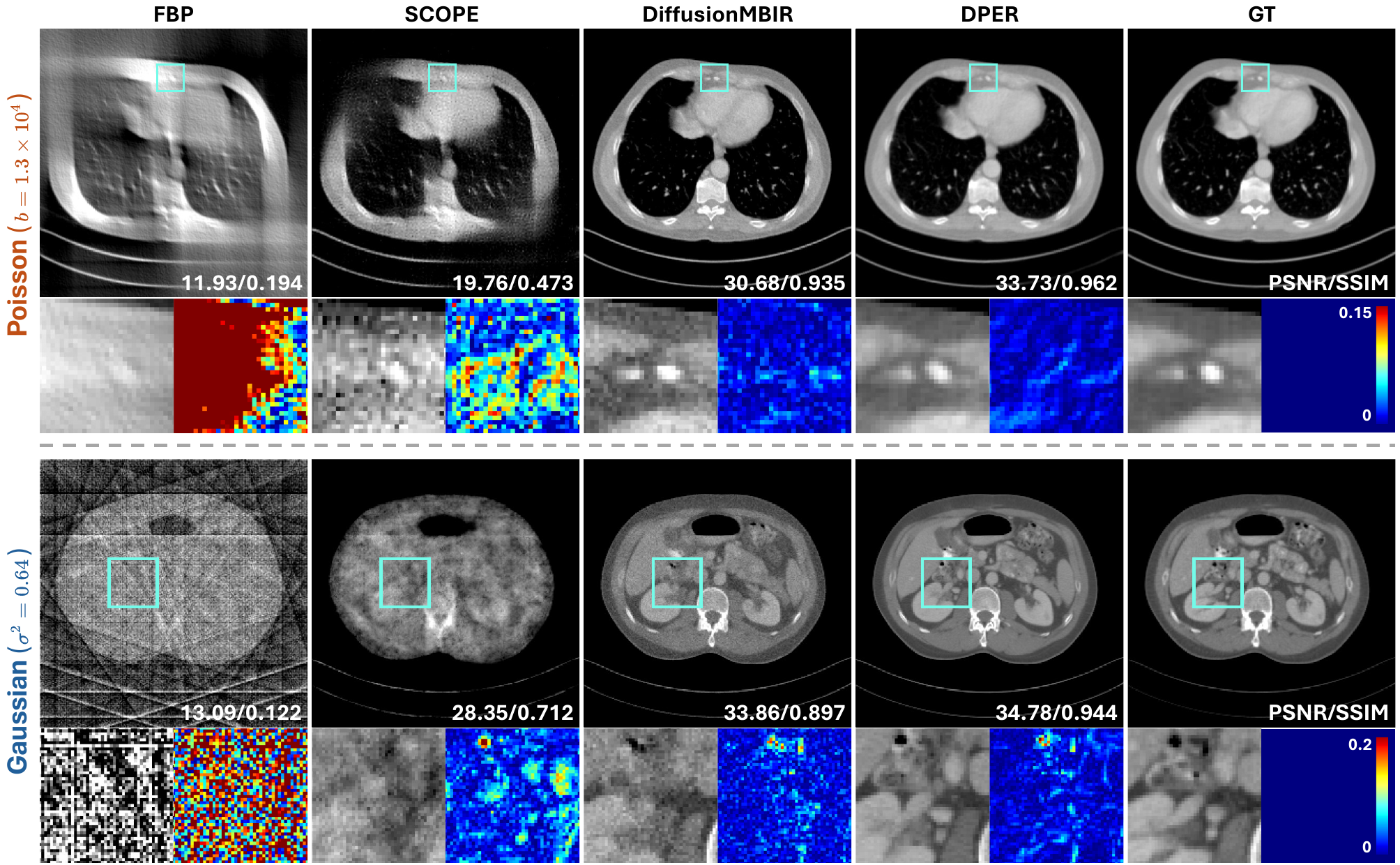}}
	\caption{{Qualitative and quantitative results of different methods to compare the measurement with Poisson noise ($b=1.3\times10^4$) and Gaussian noise ($\sigma^2=0.64$) on one test sample (\#18) of the AAPM dataset for the LACT task of [0, 90]° scanning range and SVCT task of 20 views.}}
	\label{fig:Exp_ABL_noise}
\end{figure}

Fig. \ref{fig:Exp_ABL_noise} and Table \ref{tab: different noise result} display the quantitative and qualitative results of the impact of different noise levels {and types} on our framework and other baselines for observing the robustness. There are some observations: 1) The results indicate that noise affects the model's performance as it leads to a more ill-posed LACT inverse imaging problem. It is evident that as the noise level increases, the performance of all models deteriorates. However, the reduction ratio of data-specific models shows better results than Prior-based methods, which is expected (The effect of diffusion prior is limited by noise measurement). 2) Compared to the FBP result with noise, the DPER result is significantly cleaner and maintains the data fidelity, which represents that the DPER framework can adapt to a single instance with different noise levels {and types}. 3) {Compared to the SOTA DiffusionMBIR, our framework maintains an improvement of nearly +2 dB PSNR in all noise settings for the LACT task and +1.5 dB PSNR for the SVCT task. The robustness of our method to noise is further demonstrated by the results in Fig. \ref{fig:Exp_ABL_noise}. The DiffusionMBIR results exhibit significant noise, resulting in blurred edges. In contrast, DPER stands out through effective noise reduction and clear structural details.}

To summarize, the performance of all the methods is limited by noise in the measurement domain and our DPER can still preserve the best performance in all cases.

\begin{table}

\centering
\caption{Quantitative results (PSNR / SSIM) of different methods on the AAPM dataset test samples for the LACT task with a [0, 90]° scanning range and the SVCT task with 20 views under different noise conditions (Poisson noise with SNR = 32 dB, 37 dB, 43 dB, and Gaussian noise with \(\sigma^2 = 0.64\)). The best performances are highlighted in \textbf{bold}.}
\label{tab: different noise result}
\resizebox{\columnwidth}{!}{

\begin{tabular}{cccccc} 
\toprule
\multirow{2}{*}{\textbf{Task}} & \multirow{2}{*}{\textbf{Method}} & \multicolumn{3}{c}{\textbf{Poisson Noise }}                                                            & \textbf{Gaussian Noise }  \\ 
\cmidrule(r){3-5}\cmidrule(lr){6-6}
                               &                                  & \textbf{32 dB}                   & \textbf{37 dB}                   & \textbf{43 dB}                   & $\sigma^2=0.64$           \\ 
\midrule
\multirow{4}{*}{\begin{tabular}[c]{@{}c@{}}\textbf{LACT}\\{[0,90]°}\end{tabular}}       & FBP\cite{fbp}                              & 11.43 / 0.2018                   & 11.47 / 0.2366                   & 11.51 / 0.2974                   & 12.36 / 0.2565            \\
                               & SCOPE\cite{wu2022self}                            & 22.75 /~0.6829                   & 23.08 / 0.7409                   & 23.32~/ 0.8124                   & 21.38 / 0.6867            \\
                               & DiffusionMBIR\cite{chung2023solving}                   & 31.88 /~0.9234                   & 33.07 / 0.9390                   & 34.15~/ 0.9503                   & 31.65 / 0.9210            \\
                               & DPER (\textit{Ours})             & \textbf{33.36 /}~\textbf{0.9459} & \textbf{34.52}~/~\textbf{0.9602} & \textbf{36.41}~/~\textbf{0.9685} & \textbf{33.29 / 0.9584}   \\ 
\midrule
\multirow{4}{*}{\begin{tabular}[c]{@{}c@{}}\textbf{SVCT}\\{20 \textit{views}}\end{tabular}}       & FBP\cite{fbp}                              & 15.82 / 0.1961                   & 17.62 / 0.2711                   & 20.48 / 0.4473                   & 17.12 / 0.2367            \\
                               & SCOPE\cite{wu2022self}                            & 27.73 / 0.7758                   & 28.31 / 0.8015                   & 28.24 / 0.8152                   & 27.81 / 0.7912            \\
                               & DiffusionMBIR\cite{chung2023solving}                    & 35.01 / 0.8915                   & 36.92 / 0.9280                   & 38.97 / 0.9497                   & 36.01 / 0.9072            \\
                               & DPER (\textit{Ours})             & \textbf{36.75 / 0.9517}          & \textbf{37.99 / 0.9567}          & \textbf{39.77 / 0.9661}          & \textbf{37.45 / 0.9537}   \\
\bottomrule
\end{tabular}

}

\end{table}

\section{Conclusion}
\label{sec:conclusion}
\par This paper proposes DPER, an unsupervised DL model for highly ill-posed LACT and ultra-SVCT tasks. The proposed DPER follows the HQS framework to decompose the classical CT inverse problem into a data fidelity sub-problem and a distribution prior sub-problem. Then, the two sub-problems are respectively addressed by INR and a pre-trained diffusion model. The INR guarantees data consistency with incomplete CT measurements benefiting from the integration of the CT physical forward model, while the diffusion model provides a generative image prior effectively constraining the solution space. The empirical studies on comprehensive comparison experiments reveal that the proposed DPER significantly excels both quantitatively and qualitatively. It  establishes itself as the state-of-the-art performer on in-domain and out-of-domain datasets for both the LACT and ultra-SVCT problems.

\section{Acknowledgments}
This work was supported by the National Natural Science Foundation of China (NSFC) under Grant 62071299 and Grant 12074258.

\bibliographystyle{IEEEtran}
\bibliography{refs}

\end{document}